\documentclass[twocolumn,prx,longbibliography,superscriptaddress]{revtex4-1}
\usepackage[bookmarks=false,linkcolor=blue,urlcolor=blue,colorlinks,citecolor=blue]{hyperref}
\usepackage{graphicx}
\usepackage{color}
\usepackage{amssymb}
\usepackage{amsmath}
\usepackage{dsfont}
\usepackage{bm}
\usepackage{bbm}
\usepackage{verbatim}

\newcommand{\Ket}[1]{\vert  #1 \rangle}
\newcommand{\Bra}[1]{\langle  #1 \vert}
\newcommand{\MatEl}[3]{\langle  #1 \vert \, #2 \vert #3\rangle}
\newcommand{\Amp}[2]{\langle  #1 \vert  #2 \rangle}

\newcommand{\Avg}[1]{\langle  #1  \rangle}

\newcommand{\be}{\begin{equation}}
\newcommand{\ee}{\end{equation}}
\newcommand{\bea}{\begin{eqnarray}}
\newcommand{\eea}{\end{eqnarray}}

\let\straightphi\phi

\renewcommand{\phi}{\varphi}

\renewcommand{\epsilon}{\varepsilon}

\begin{document}

\title{Universal chiral quasi-steady states in periodically driven many-body systems}

\author{Netanel H. Lindner}
\affiliation{Physics Department, Technion, 320003 Haifa, Israel}
\author{Erez Berg}
\affiliation{Department of Condensed Matter Physics, The Weizmann Institute of Science, Rehovot, 76100, Israel}
\author{Mark S. Rudner}
\affiliation{Niels Bohr International Academy and Center for Quantum Devices, University of Copenhagen, 2100 Copenhagen, Denmark}

\date{\today}

\begin{abstract}
We investigate many-body dynamics in a one-dimensional interacting periodically driven system, based on a partially-filled version of Thouless's topologically quantized adiabatic pump.
The corresponding single particle Floquet bands are {\it chiral}, with the Floquet spectrum realizing nontrivial cycles around the quasienergy Brillouin zone. 
For generic filling, with either bosons or fermions, the system is gapless 
and is expected to rapidly absorb energy from the driving field. 
We identify parameter regimes where scattering between Floquet bands of opposite chirality is exponentially suppressed, opening a long time window in which the system prethermalizes to an infinite temperature state restricted to a single Floquet band. 
In this quasi-steady state, 
the time-averaged current takes a universal value determined solely by the density of particles and the topological winding number of the populated Floquet band.
This remarkable behavior 
may be readily studied experimentally in recently developed cold atom systems. 
\end{abstract}

\maketitle

\section{Introduction}
Topological transport has 
garnered great attention in condensed matter physics, ever since 
the discovery of the quantized Hall effects \cite{vonKlitzing1980,Laughlin1981}.
Traditionally, robust features of transport have been linked to topological properties of the ground states of many-body systems. 
\begin{figure}[h!t]
\includegraphics[width=0.950 \columnwidth]{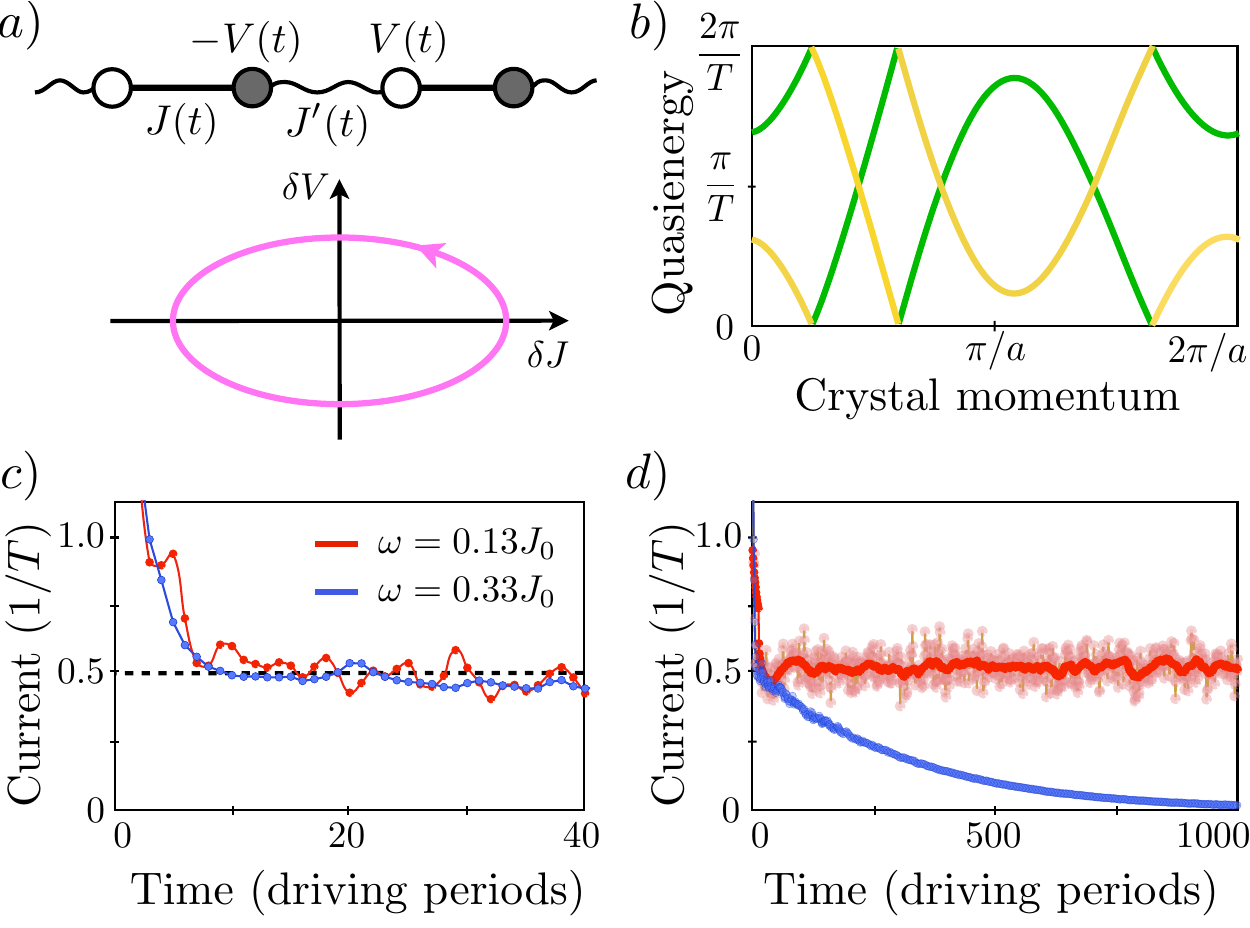}
\caption{Establishing the universal chiral quasi-steady state in a one-dimensional periodically driven system. 
a) Tight binding model, see 
Eq.~(\ref{eq: h0}). The dimerized hopping amplitudes and on-site potentials are changed adiabatically as depicted by the magenta curve. The origin $\delta J=\delta V=0$ is a degeneracy point.
b) The Floquet spectrum for $\omega =0.2J_0$ and $\lambda=1$, {see text below Eq.~(\ref{eq: h0})}. The right (left) moving Floquet band is colored green (yellow).
c) The period-averaged current vs.~time obtained from a numerical simulation of a system of length $L=16$ unit cells with $N=8$ particles, for two different driving frequencies (see legend).
After a few driving periods the system reaches a quasi-steady state featuring a current $\mathcal{J} \approx \rho/T$, 
where $\rho=N/L$ is the density of particles. 
d) At low frequency ($\omega = 0.13 J_0$, red) the chiral quasi-steady state persists for thousands of periods without noticeable decay.
Interband scattering leads to a decay of the current with a rate that falls exponentially with $1/\omega$ (visible here for $\omega = 0.33 J_0$, blue).
In (c) and (d): $\lambda=0.66$, $U=2J_0$.}
\label{fig:summary}
\vspace{-0.2 in}
\end{figure}
Recently, a new paradigm has emerged for altering 
the topological properties of band structures using external driving 
\cite{Oka2009,Inoue2010,KBRD,Lindner2011,Lindner2013, Gu11,Kitagawa2011,Delplace2013,Podolsky2013,Liu2013,Titum2015,TorresPRB2014,TorresPRL2014,AlessioArxiv2014, Dehghani2014, Dehghani2014b, Sentef2015,Seetharam2015, Iadecola2015, Klinovaja2015, Gannot2015}.
These ideas have sparked a variety of proposals and initial experiments aimed at realizing so-called
Floquet topological insulators in a range of solid state, atomic, and optical systems \cite{Wang2013, Gedik2016, Jotzu2014,Rechtsman2013, Hu2015}.
While the prospect of dynamically controlling band topology is 
exciting, establishing how and to what extent topological phenomena can be observed in such systems raises many crucial and fundamental questions about many-body dynamics in periodically driven systems.

The long-time behavior of periodically-driven many-body systems is a fascinating question of current investigation.
Several recent theoretical works 
suggest that closed, interacting, driven many-body systems 
generically absorb energy indefinitely from the driving field, tending to infinite-temperature-like states in the long time limit~\cite{LazaridesDasMoessner2014,DAlessio2014,Lazarides2015}.
In such a state, all correlations are trivial, indicating in particular that any
topological features of the underlying Floquet spectrum are expected to be washed out. 
On the other hand, several interesting exceptions to the infinite temperature fate have been proposed~\cite{Citro2015, Chandran2015, Prozen}.
Particularly, heating may be circumvented by local conservation laws, e.g., 
in integrable~\cite{Lazarides2014} or many-body localized systems~\cite{Ponte2014, Abanin2014, Khemani2015, vonKeyserlingk2016a, vonKeyserlingk2016b, Potter2016, Nayak2016, Roy2016}.

In this work we describe an intriguing new possibility: rather than treating heating as an enemy, we show that it can be used as a resource for establishing universality and a novel regime of topological transport in interacting periodically driven systems.
The universal ``quasi-steady'' state that develops persists for an {\it exponentially long} time, controlled by the inverse driving frequency.
At very late times the state eventually
gives way to a featureless infinite temperature state as described above.

Previously, metastability and exponentially slow heating rates have been discussed for the limit of high frequency driving~\cite{AbaninHighFreq1,AbaninHighFreq2, Eckardt2015, Bukov2015},
where the driving frequency is much larger than all local energy scales.
In this case, absorption can only occur through high-order multiparticle processes.
The system then ``prethermalizes'' to an equilibrium-like state governed by a local, static, effective Hamiltonian~\cite{Berges2004,Eckstein2009,Moeckel2010,Mathey2010,BukovPrethermal, Bukov2015, Eckstein2015, Kuwahara2016, Mori2015}.
In this work we introduce a qualitatively different regime of prethermalization, which occurs for {\it low frequency} driving in systems with well-separated high and low energy degrees of freedom.
Rapid heating within the restricted low-energy subspace plays an essential role in  establishing the universal quasi-steady state, while exponentially-suppressed heating rates of the high energy degrees of freedom to preserve this state. 
The exponential separation of timescales between equilibration and decay allows the quasi-steady regime to be treated as a stable phase for practical purposes.


We focus our attention to one dimension (1d), where Thouless showed that cyclic adiabatic driving in filled-band/gapped many-body systems leads to topologically-quantized charge pumping~\cite{Thouless1983}. 
This phenomenon was recently observed in cold atoms experiments~\cite{lohse2015thouless, nakajima2016topological}.
The average current carried by the system over one driving period is quantized when one of the Floquet bands of this periodically driven lattice system is completely filled with fermions, and the other is completely empty.
What happens when the bands are initially only \emph{partially} filled, or if the system is comprised of bosons? Here the system lacks a many-body gap, and 
the evolution cannot be considered to be adiabatic. In the absence of interactions any value of the pumped charge per cycle is possible, depending on the details of the Hamiltonian and the initial state. However, in the interacting case we identify a parameter regime where heating naturally drives the system to a quasi-steady state with universal properties, {reflecting} 
the topological nature of the Floquet band structure.

\section{Summary of main results}
We now briefly describe the setup and our main results, giving details in the sections that follow.
The Thouless pump is exemplified by a time-dependent tight-binding model, as illustrated in Fig.~\ref{fig:summary}a [see Eq.~(\ref{eq: h0}) below].
In the adiabatic limit, the system's single-particle Floquet spectrum exhibits a {\it non-trivial winding} of each band~\cite{KBRD}: the quasi energy changes by $\omega \equiv 2\pi/T$ (where $T$ is the driving period) as the {crystal momentum changes from $0$ to $2\pi/a$, where $a$ is the lattice constant}~\cite{KBRD}\footnote{For any finite drive frequency $\omega$, minigaps open at the crossing between the Floquet bands. These minigaps are suppressed exponentially in $1/\omega$.}. A characteristic Floquet spectrum with such winding (or ``chiral'') bands is shown in Fig.~\ref{fig:summary}b.

We study dynamics in the situation where one of the chiral Floquet bands is initially partially filled, while the other is empty. 
We assume that the minimum gap between the two bands of the single particle 
Hamiltonian (minimized over all times and quasi-momenta), $\Delta$, is much larger than $\mathrm{max}[\hbar \omega,U,W]$,  where $U$ and $W$ are the interaction strength and the maximum bandwidth, respectively (see below for a precise definition of $W$).
Based on an analysis of low and high order scattering rates, we argue that that (i)
the system equilibrates on a short time scale $\tau_{\rm intra}$
to a \textit{chiral infinite-temperature-like state} in which all momentum states in one of the chiral bands are equally populated, while the other band remains nearly empty. This quasi-steady state is obtained regardless of the initial state, as long as only one Floquet band is initially populated; (ii) In this state, the average pumped charge per period is approximately $w \rho$, where $w$ is the winding number of the partially filled band (see Sec.~\ref{sec: Model})  and $\rho$ is the density of particles~\footnote{A conservative estimate of the corrections to this value of the current is $\mathrm{O}[(U/\Delta)^2]$, due to virtual transitions between the bands.}; and (iii) the quasi-steady state persists for a long time scale $\tau_{\rm inter}$, that can be larger than $\tau_{\rm intra}$ by many orders of magnitude.
In particular, for a broad parameter regime where $\Delta\gg\omega\gtrsim W$, we find that $\tau_{\mathrm{intra}}/\tau_{\mathrm{inter}} \propto (U/\Delta)^{b \Delta/\omega}$, where $b$ is a number of order unity.
At times $t \gg \tau_{\mathrm{inter}}$, the system relaxes to a state in which the values of all local observables are as in an infinite temperature state. In particular, the  current tends to zero.
This physical picture is supported by exact numerical simulations of finite systems, see Figs.~\ref{fig:summary}c,d.

What is the nature of the novel quasi-steady state exhibited by the system? 
We consider {\it unitary} dynamics of a closed, driven system.
Therefore a pure initial state remains pure for all times.
Moreover, strictly speaking, the system does not possess a steady state.
Consider, however, the behavior of generic local observables (in the spirit of previous works on ergodicity and thermalization in non-driven systems~\cite{Deutsch91, Srednicki99, Reimann2008}).
Importantly, the expectation value of a local observable depends only on the reduced density matrix of a subsystem that contains the region where the observable is defined.
From this point of view, the system may effectively be described by a mixed (possibly thermal) state.

In the sense described above, an ergodic periodically-driven system is generically expected to tend toward an infinite temperature state.
In an infinite temperature ensemble, all microstates (many-body eigenstates) appear with equal probability, as characterized by a subsystem reduced density matrix $\hat{\rho}_\infty \propto \hat{1}$.
In our case, the {\it dynamical constraint} imposed by slow inter-band scattering allows the system to approach an infinite-temperature-like quasi-steady state {\it restricted to one band}, $\alpha$, in which all {\it available} states are equally probable: $\hat{\rho}_\mathrm{qs}(t) \propto P_\alpha(t)$, where $P_\alpha(t)$ is a (time-dependent) projection onto all many-body states with a given number of particles in Floquet band $\alpha$.

In the band-restricted infinite temperature state, the system synchronizes with the drive; observables may oscillate over the course of one driving period, but their values remain constant from one period to the next.
In particular, the state features equal occupation numbers of all Floquet modes in band $\alpha$, $N_\alpha(k) = {\rm const}$.
The loss of memory of the initial occupation distribution allows the universal (topological) characteristics of the Floquet bands to be exposed by the dynamics.

As we shall see, our numerical findings for the quasi-steady state are consistent with the form of the density matrix described above. In particular, we can readily compute the expectation value of the current in this state.
While the current $\hat{J}(t)$ exhibits non-universal oscillations, its time average over one period is simple and universal: $\mathcal{J}   = \int_{t}^{t+T}dt'\,\mathrm{Tr}[\hat{J}(t') \hat{\rho}_{\mathrm{qs}}(t')]/\mathrm{Tr}[\hat{\rho}_{\mathrm{qs}}(t')] = \rho w/T$, where $\tau_{\rm intra} \ll t \ll \tau_{\rm inter}$ is any time after the quasi-steady state has been reached. We confirm this universal behavior for a variety of parameter choices in both trivial $(w=0)$ and nontrivial $(w=1)$ topological phases, see Sec.~\ref{sec: Numerics}.
For asymptotically long times, $t \gg \tau_{\rm inter}$, the system tends to 
an {\it unrestricted} infinite temperature steady state. 
In this state, \emph{the average current vanishes at all times}, $\Avg{\mathcal{J}(t)} = 0$, for any driving.

\section{Model}
\label{sec: Model}
We begin our analysis by defining the specific model that we use to illustrate the universal chiral quasi-steady regime. 
We consider a 1d lattice with two sites per unit cell, labeled $A$ and $B$. The hopping matrix elements and on-site potentials are time dependent (see Fig.~\ref{fig:summary}a). The lattice is populated by a finite density $\rho$ of identical fermions or bosons per unit cell. 

We write the Hamiltonian as $H(t) = H_0(t) + H_{\mathrm{int}}$, where the single particle part $H_0(t)$ is given by
\begin{eqnarray}
H_0(t) &=& -J(t) \sum_{j} \hat{c}^{\dagger}_{j,A} \hat{c}^{\vphantom{\dagger}}_{j,B} - J'(t)\sum_{j} \hat{c}^{\dagger}_{j,B} \hat{c}^{\vphantom{\dagger}}_{j+1,A} + {\rm h.c.} \nonumber\\
&+& V(t) \sum_j (\hat{c}^\dagger_{j,A}\hat{c}^{\vphantom{\dagger}}_{j,A} - \hat{c}^\dagger_{j,B}\hat{c}^{\vphantom{\dagger}}_{j,B}).
\label{eq: h0}
\end{eqnarray}
Here, $\hat{c}^\dagger_{j,A}$ ($\hat{c}^\dagger_{j,B}$) creates a particle of type $A$ ($B$) in unit cell $j$.
To avoid later ambiguity, we use hats to denote creation and annihilation operators throughout this work.
Below we take $J(t) = J_0 + \delta J(t)$, $J'(t) = J_0 - \delta J(t)$, and $V(t) = V_0 + \delta V(t)$, with $\delta J(t) = \delta J_1 + \delta J_2 \cos\omega t$ and $\delta V(t) = V_1 \sin\omega t$.
{For convenience, we parametrize $\delta J_2 = \lambda J_0$ and $V_1 = 3\lambda J_0$ in most of the simulations presented below.}

%
We consider a local interaction 
\begin{equation}
H_{\mathrm{int}}= \tfrac{1}{2} U \sum_{j} n_{j}(n_{j}-1),\ n_{j} = \hat{c}^\dagger_{j,A}\hat{c}^{\vphantom{\dagger}}_{j,A} + \hat{c}^\dagger_{j,B}\hat{c}^{\vphantom{\dagger}}_{j,B}.
\label{eq: hint}
\end{equation}
The intra-unit-cell form of the interaction in Eq.~(\ref{eq: hint}) 
{is convenient for the analysis below.}
However, we do not expect our conclusions to depend on this choice. 

The single particle Floquet spectrum corresponding to $H_0(t)$ is found by seeking solutions to the Schr\"{o}dinger equation which satisfy~\footnote{We take $\hbar = 1$ throughout.}: 
$\Ket{\Psi_{\rm 1P}(t)} = e^{-i\epsilon_{\rm 1P} t}\Ket{\Phi_{\rm 1P}(t)}$, with $\Ket{\Phi_{\rm 1P}(t + T)} = \Ket{\Phi_{\rm 1P}(t)}$.
We decompose the periodic function $\Ket{\Phi_{\rm 1P}(t)}$ in terms of an infinite set of (non-normalized) discrete Fourier modes $\{\Ket{\phi_{\rm 1P}^{(m)}}\}$:
\be
\label{eq:FloquetState} \Ket{\Psi_{\rm 1P}(t)} = e^{-i\epsilon_{\rm 1P} t}\sum_m \Ket{\phi_{\rm 1P}^{(m)}}e^{-im\omega t}.
\ee
The full time-dependent evolution of $\Ket{\Psi_{\rm 1P}(t)}$ is specified by the quasienergy $\epsilon_{\rm 1P}$ and a vector of Fourier coefficients
 \be
\label{eq:phi_vector} \phi_{\rm 1P} =
 \left(
   \begin{array}{c}
     \vdots\\
      \Ket{\phi_{\rm 1P}^{(-1)}}\\
      \Ket{\phi_{\rm 1P}^{(0)}}\\
      \Ket{\phi_{\rm 1P}^{(1)}}\\
      \vdots
   \end{array}
 \right).
 \ee
Throughout this work we choose the quasi-energies of all single particle states to lie within a fundamental Floquet-Brillouin zone, $0 \le \epsilon_{\rm 1P} < 2\pi/T$.

The Fourier coefficients comprising $\Ket{\Psi(t)}$ are determined by an 
eigenvalue equation $\epsilon\phi = \mathcal{H}_0 \phi$, where the ``extended Hamiltonian'' $\mathcal{H}_0$ is constructed from the Fourier decomposition of $H_0(t)$ (see Appendix~\ref{sec: single particle}) and $\phi$ without a ket symbol stands for a column vector of Fourier modes as in Eq.~(\ref{eq:phi_vector}). We use calligraphic symbols for
matrices in the space of Fourier coefficients.
For harmonic driving, $H_0(t) = H_{\rm dc} + \Lambda e^{i\omega t} + \Lambda ^\dagger e^{-i\omega t}$, the matrix $\mathcal{H}$ takes a simple block tri-diagonal form in harmonic ($m$) space: $(\mathcal{H}_0)_{mm'} = (H_{\rm dc} + m\omega)\delta_{mm'} + (\Lambda\delta_{m,m'-1} + \Lambda^\dagger \delta_{m,m'+1})$. 
The single particle Floquet spectrum is shown in Fig.~\ref{fig:summary}b for parameters specified in the caption.
For each of the two bands, we assign a winding number $w$, such that the quasi-energy band winds  by $w\omega$ when the quasi-momentum $k$  changes from $0$ to $2\pi/a$. In our case, the two bands have winding numbers $w=+1$ and $w=-1$, and we refer to them as the right-moving (R) and left-moving (L) bands, respectively. More generally, a non-trivial winding is achieved in the adiabatic limit when the curve $(\delta J(t), \delta V(t) )$ encircles the origin (as in Fig.~\ref{fig:summary}a).

{
\section{Many-body dynamics}
We now turn to the many-body dynamics of this system.
We consider the situation where the system is initialized with a finite density of particles in the net right-moving (R) Floquet band, shown in green in Fig.~\ref{fig:summary}b.
The initial momenta of the particles are arbitrary. 
}

To investigate the timescales for intra-band and inter-band scattering, 
we develop a perturbative analysis of 
the many-body dynamics of the system.
The perturbation series is organized in terms of powers of the interaction strength $U$.
Crucially, we work in a Floquet picture where the time-dependent driving is first taken into account exactly, to all orders in the driving.
As above, we work in the extended space of Fourier coefficients, 
where the many-body Floquet eigenstates are described by the eigenvectors of the extended Hamiltonian, $\mathcal{H} = \mathcal{H}_0 + \mathcal{U}$, with $\mathcal{U}_{mm'} = H_{\rm int}\delta_{mm'}$.

{
The extended Hamiltonian {defines a} (static) eigenvalue problem that yields the Fourier coefficients describing Floquet eigenstates.
One may also use the extended Hamiltonian to generate an effective evolution in the extended space, in an auxiliary time variable $\tau$, via $i \partial_\tau  \phi(\tau) = \mathcal{H}\phi(\tau)$.
For the stroboscopic times $\tau = nT$, where $n$ is an integer and $T$ is the driving period of the original problem, the ``evolved'' vector of Fourier coefficients $\phi(nT)$ precisely captures the state of the system in the physical Hilbert space at the corresponding time $t = nT$ (see Appendix~\ref{sec: stroboscopic}).
Using this mapping we obtain transition rates for the stroboscopic evolution by employing standard Green's function techniques to the auxiliary evolution problem in the extended space.
}

{
Our aim is to calculate the rate at which particles are scattered into the left-moving (L) band.
For weak interactions and short times, it is natural to view this process in terms of a perturbation series in the interaction $\mathcal{U}$.
We express the auxiliary-time evolution of the Fourier vector $\phi(\tau)$ in terms of its Fourier transform, $\phi(\tau) = \int_{-\infty}^\infty d\tau\,e^{i\Omega\tau} \tilde{\phi}(\Omega)$.
In terms of the extended Green's function $\mathcal{G}_0(\Omega) = (\Omega - \mathcal{H}_0 + i\delta)^{-1}$ and T-matrix $\mathcal{T}(\Omega) = \mathcal{U} + \mathcal{U}\mathcal{G}_0(\Omega)\mathcal{T}(\Omega)$, we have
\be
\tilde{\phi}(\Omega) = i\,[\mathcal{G}_0(\Omega) + \mathcal{G}_0(\Omega)\mathcal{T}(\Omega)\mathcal{G}_0(\Omega)]\,\phi_0,
\ee
where $\phi_0$ is the Fourier vector corresponding to the ``free'' initial state in which all particles are initialized in single-particle Floquet eigenstates in the right-moving (R) band of the non-interacting system (see Appendix \ref{appendix:FreeStates} for details of the construction of $\phi_0$).
}

{
\subsection{Born approximation}As a first step, we investigate the scattering rates to leading order in $\mathcal{U}$, i.e., in the Born approximation $\mathcal{T}(\Omega) \approx \mathcal{U}$. 
This approximation captures the leading-order behavior of two-particle scattering, which one may expect to be relevant for weak interactions and low densities (see discussion below {and Refs.~\cite{Choudhury2015, Bilitewski2015, Genske2015}}). 


Within the Born approximation, the transition rate is given by
$\Gamma \approx 2\pi \sum_{f \ne 0} {\delta(E_f - E_0)} | \varphi_f^\dagger\, \mathcal{U} \varphi_0 |^2$.
This is Fermi's golden rule, adapted for a Floquet system~\footnote{{The Fermi's golden rule rates are evaluated using ``free'' initial and final states, as taken at $t = 0$. For $t \gtrsim \tau_{\rm intra}$, the state locally appears similar to an infinite temperature state (within the R band), which can be described as a uniform mixture of all free states in the band.}}. Here $f$ labels all 
final {``free''} Floquet eigenstates, and $\{E_f\}$ are their eigenenergies (with respect to the extended Hamiltonian $\mathcal{H}_0$). We break $\Gamma$ into two parts, $\Gamma = \Gamma_{\mathrm{intra}} + \Gamma_{\mathrm{inter}}$, corresponding to intra-band and inter-band scattering, respectively. The latter scattering processes {transfer} 
one or more particles from the R to the L band.

Figures~\ref{fig:TwoParticleScattering}a,b show the two-particle scattering rates for a pair of {bosons} in the R band, with momenta $\{k_1, k_2 \}$, to scatter either to states within the R band (processes we denote as RR$\rightarrow$RR, Fig.~\ref{fig:TwoParticleScattering}a), or to states with one particle in the R band and one in the L band (RR$\rightarrow$RL, Fig.~\ref{fig:TwoParticleScattering}b).  The two-particle scattering rates are given by 
\be
\frac{\Gamma_{\mathrm{2P}}}{L} = \sum_f \frac{1}{|\Delta v_f|} | \phi_{\mathrm{2P}, f}^\dagger\, \mathcal{U} \phi^{\vphantom{\dagger}}_{\mathrm{2P},0} |^2,
\label{eq: golden}
\ee
where $\phi_{\mathrm{2P},0}$ and $\phi_{\mathrm{2P},f}$ are the  wave functions of the initial and final two-particle Floquet states (see Appendix \ref{appendix:FreeStates}), $\Delta v_f$ is the difference of group velocities of the two outgoing particles, and the summation is over all final states that satisfy quasi-energy and quasi-momentum conservation. The factor {$1/|\Delta v_f|$} comes from the density of outgoing states.

\begin{figure}[t]
\includegraphics[width=1.0 \columnwidth]{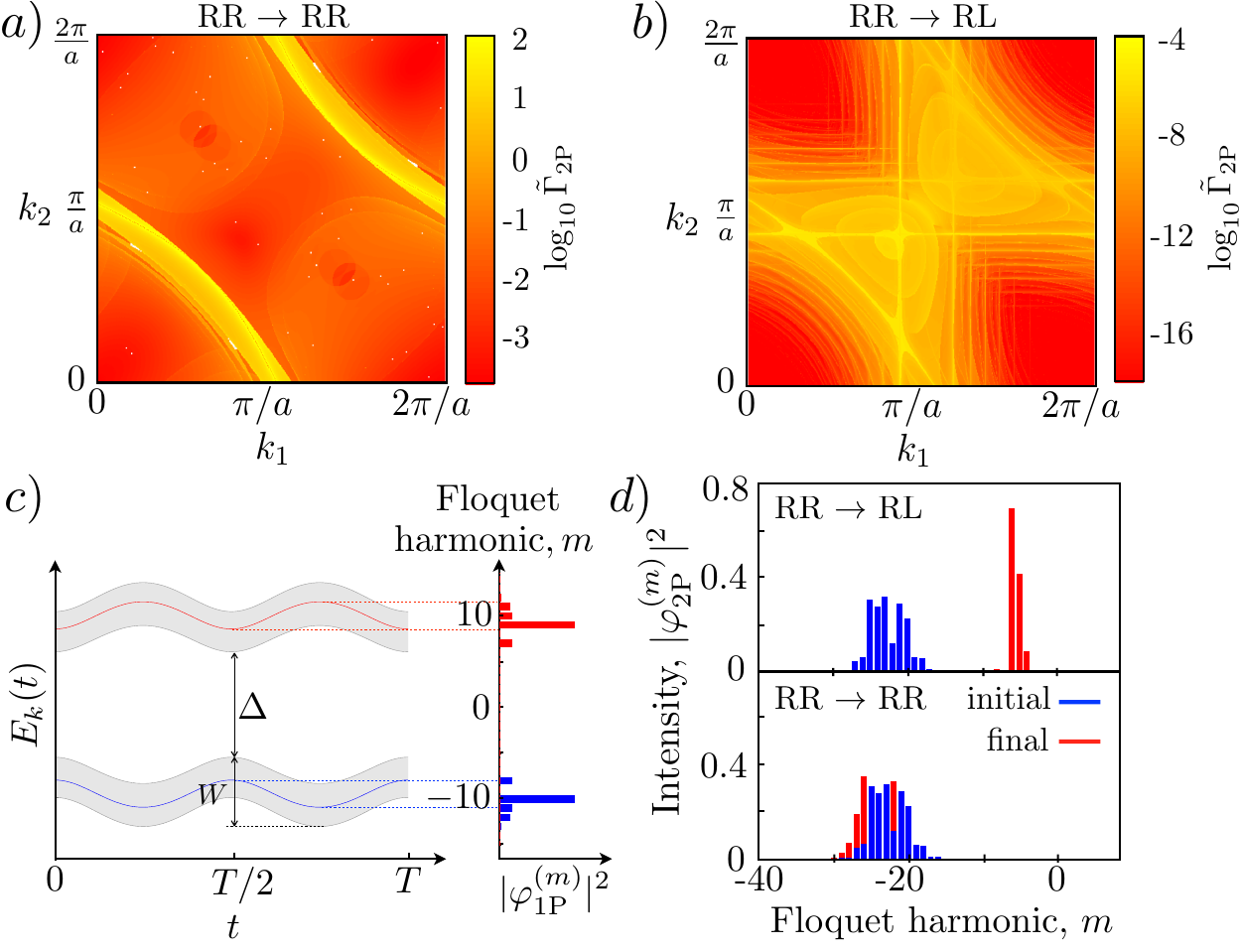}
\caption{Two-particle scattering in a partially-filled Thouless pump.
a) Intraband scattering rate $\tilde{\Gamma}_{\rm 2P} = \Gamma_{\rm 2P}/(J_0L)$, evaluated {for bosons} within Fermi's golden rule, Eq.~(\ref{eq: golden}), with forward scattering contribution removed. 
Here $k_1$ and $k_2$ are the momenta of two incident particles, and we take $\lambda = 0.66$, $\omega = 0.13J_0$, and $U = 0.67 J_0$.
b) Same as above, for interband scattering in which one particle is scattered from the right to the left moving band.
c) Single particle Floquet states. On the left, the shaded region spans the energies of the instantaneous bands as a function of time. The significant Fourier components $|\phi_{\rm 1P}^{(m)}|^2=|\Bra{\phi_{\rm 1P}^{(m)}}\phi_{\rm 1P}^{(m)}\rangle|^2$ of a Floquet state $\alpha$ with momentum $k$ fall between the minimum and maximum values of the instantaneous energies $E_{\alpha,k}(t)$. 
d) Fourier components $|\phi_{\rm 2P}^{(m)}|^2=|\Bra{\phi_{\rm 2P}^{(m)}}\phi_{\rm 2P}^{(m)}\rangle|^2$ of incoming and outgoing two particle Floquet states. For intraband scattering (bottom) the Fourier components overlap. For interband scattering (top), the overlap is strongly suppressed leading to a suppression of the interband scattering rate.}
\label{fig:TwoParticleScattering}
\end{figure}


Our key observation is that \emph{the inter-band scattering rates are strongly suppressed compared with the intra-band ones.} For the parameters chosen, the mean $\textrm{RR}\rightarrow\textrm{RL}$ scattering rate is down by a factor $\sim 10^{-8}$ 
compared to the average $\textrm{RR}\rightarrow\textrm{RR}$ rate. The average $\textrm{RR}\rightarrow\textrm{LL}$ rate (not shown) is suppressed by a factor of $\sim 10^{-9}$ relative to $\textrm{RR}\rightarrow\textrm{RL}$. {The bright lines visible in the interband scattering rate, Fig.~\ref{fig:TwoParticleScattering}b, are associated with single particle resonances; such resonances yield significant interband scattering rates only in exponentially small regions of phase space (see Appendix~\ref{sec: single particle} for further discussion of the role of these single-particle resonances).}

The suppression of inter-band relative to intra-band scattering originates from the matrix element in Eq.~(\ref{eq: golden}).
Since $\mathcal{U}$ is diagonal in Fourier harmonics, the matrix element for inter-band scattering is suppressed if the initial and final states have support in different regions in harmonic space.
In the adiabatic limit $\omega \ll \Delta$, this is indeed the case. Figure~\ref{fig:TwoParticleScattering}c shows two representative single-particle Floquet wave functions with quasi-momenta $k_1$, $k_2$, one from each band. The support of each state in harmonic space corresponds to the energy window spanned by the \emph{instantaneous energy}, $E_{\alpha,k}(t)$ (the eigenvalue of $H_0(t)$ for band $\alpha=\textrm{R,L}$), with $0 \le t < T$~\cite{comment-quasienergy}. This energy window is bounded by $W$, which we define as $W = \mathrm{max}_{k,t}{E_{R,k}(t)} -  \mathrm{min}_{k,t}{E_{R,k}(t)}$, see Fig.~\ref{fig:TwoParticleScattering}c. Outside of this window, the Floquet wave functions decay rapidly. 
The separation of the Floquet states of the two bands in harmonic space can be derived by mapping the Floquet problem to a Zener tunneling problem 
in a weak electric field (see Appendix~\ref{sec: single particle}). 

Figure \ref{fig:TwoParticleScattering}d shows representative two-particle states that participate in either inter- or intra-band scattering. 
The two-particle states are constructed as convolutions of two single-particle states. For intra-band scattering ($\textrm{RR} \rightarrow \textrm{RR}$), the initial and final states occupy the same region in harmonic space. In contrast, for inter-band scattering ($\textrm{RR} \rightarrow \textrm{RL}$) the initial and final states are separated in harmonic space by a gap of order $\Delta/ \omega$. (Note that this requires the energy spread of the single particle states, of order $W$, to be smaller than $\Delta$.) Hence, at least within the Born approximation, inter-band scattering is strongly suppressed with respect to intra-band scattering.

\subsection{Higher order contributions} Next, we consider higher order contributions to the inter-band scattering rate~\cite{doublon}.
Importantly, the strong suppression of the Born-level interband scattering rate arises from the exponentially small overlap of harmonic-space wave functions of the initial and final states, along with the fact that the time-independent interaction conserves the harmonic index.
As we now show, this exponential factor may be avoided at higher orders in perturbation theory, trading the small matrix element for higher powers of the interaction, $U$.
Optimizing over the competition between small $U$ and small overlaps, we find an optimal order $N_{\rm min}$ at which many-body resonances~\cite{Bukov2015} dominate 
the scattering amplitude.
In this way we argue that the scattering rate should be a power law in $U$, with an exponential suppression in the inverse of the driving frequency, $1/\omega$.

To analyze the scattering amplitudes at higher orders in the interaction, we return to the expansion of the T-matrix,

\begin{equation}
\mathcal{T}(\Omega) = \mathcal{U} + \mathcal{U}\mathcal{G}_0(\Omega)\mathcal{U}+\mathcal{U}\mathcal{G}_0(\Omega)\mathcal{U}\mathcal{G}_0(\Omega)\mathcal{U}\ + \cdots.
\end{equation} 

Recall that the operators $\mathcal{T}(\Omega)$, $\mathcal{G}_0(\Omega)$ and $\mathcal{U}$ are defined in the ``extended space'' of Fourier harmonics.
As we will see, the additional state-space dimensions of the extended space play an important role in describing energy (photon) absorption from the drive.

To facilitate the T-matrix analysis, we define a basis of non-interacting eigenstates of the extended space Hamiltonian $\mathcal{H}_0$.
For each non-interacting $N$ particle Floquet state we associate a label ${\bm \xi} = \{k_n, \alpha_n\}$, $n = 1\ldots N$,  denoting the quasimomenta and the band indices $\alpha_n = \{{\rm R}, {\rm L}\}$ of all particles.
As in the single particle case, cf.~Eq.~(\ref{eq:phi_vector}), the $N$ particle non-interacting Floquet state $|\Psi_{\bm{\xi}}(t)\rangle$ is represented by a vector of Fourier coefficients which we denote by $\phi_{\bm{\xi}}$.
We use the convention that the many-body quasi-energy is given by the sum of single particle quasi-energies, $\epsilon_{\bm \xi} = \sum_n \epsilon_{{\rm 1P}, k_n\alpha_n}$, with $0 \le \epsilon_{{\rm 1P}, k_n\alpha_n} < 2\pi/T$ as above.
Under this convention, the quasi-energies $\epsilon_{\bm \xi}$ generically fall outside of the fundamental Floquet zone.
The detailed construction of the many-body Fourier vectors $\phi_{\bm \xi}$ is given in Appendix~\ref{appendix:FreeStates}.

In the extended space representation this spectrum is repeated over and over again, shifted by integer multiples of $\omega$, with the corresponding eigenvectors likewise shifted in harmonic space. 
Therefore, with each Floquet state $\phi_{\bm{\xi}}$ defined above, we can associate a whole family of eigenstates $\{\phi_{{\bm \xi},\nu}\}$ with components $\phi_{{\bm \xi},\nu}^{(m)} = \phi_{\bm \xi}^{(m+\nu)}$ and quasienergies $\epsilon_{\bm \xi} + \nu \omega$. (Note that complete information about the physical Floquet spectrum is captured by eigenstates with a single $\nu$.) In terms of these ``copy states'' we write the extended Green's function $\mathcal{G}_0(\Omega)$ as
\be
\label{eq:G0} \mathcal{G}_0(\Omega) = \sum_{{\bm \xi}, \nu} \frac{\phi_{{\bm \xi},\nu}\phi^\dagger_{{\bm \xi},\nu}}{\Omega - \epsilon_{\bm \xi} - \nu\omega + i\delta}.
\ee
Crucially, the ``copy states'' defined above play an important role as off-shell virtual intermediate states in the perturbation theory.

At each order, the scattering amplitude involves a product of many-body matrix elements of the form $\phi^\dagger_{{\bm \xi},\nu}\mathcal{U}\phi_{{\bm \xi}',\nu'}$.
Such a matrix element can be expressed in the time-domain as
\be
\label{eq:TimeDomainMatEl} \phi^\dagger_{{\bm \xi},\nu}\mathcal{U}\phi_{{\bm \xi}',\nu'} = \frac{1}{T}\int_0^T\!\!\! dt\, e^{-i(\nu-\nu')\omega t}\MatEl{\phi_{\bm \xi}(t)}{H_{\rm int}}{\phi_{\bm \xi'}(t)},
\ee
where $\Ket{\phi_{\bm \xi}(t)} = \sum_m e^{-im\omega t}\Ket{\phi_{\bm \xi}^{(m)}}$ is the periodic part of the (non-interacting) many-body Floquet wave function.


Importantly, $H_{\rm int}$ is a two-body operator.
Therefore, computation of the matrix element in Eq.~(\ref{eq:TimeDomainMatEl}) can be reduced to the problem of evaluating matrix elements between {\it two-particle} Floquet states, $M^{(\nu-\nu')}_{\xi_1\xi_2,\xi'_1\xi'_2} = \frac{1}{T}\int_0^T\!\!\! dt\, e^{-i(\nu-\nu')\omega t}\MatEl{\phi_{{\rm 2P},\, \xi_1\xi_2}(t)}{H_{\rm int}}{\phi_{{\rm 2P},\, \xi_1'\xi'_2}(t)}$.
By further understanding the properties of these two-particle matrix elements, we may thus characterize the various terms in the high-order perturbation theory.


\begin{figure}[t]
\includegraphics[width=1.0 \columnwidth]{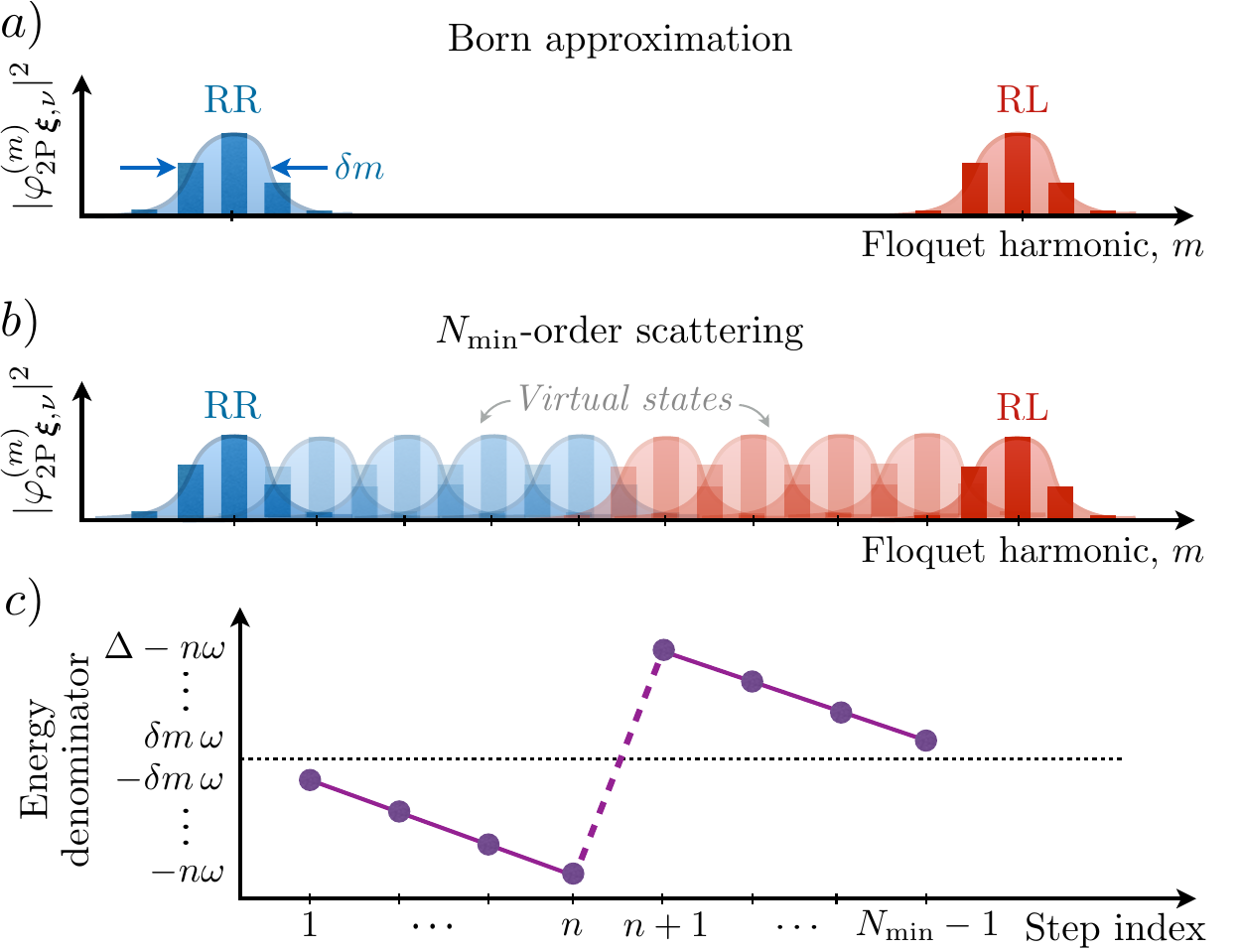}
\caption{Matrix elements and energy denominators for high-order scattering.
a) In the Born approximation, the  RR $\rightarrow$ RL scattering matrix element is (super) exponentially suppressed by the small overlap of harmonic space wave functions.
b) At an order $N_{\rm min} \sim \Delta/(\delta m \omega)$ in perturbation theory, a non-suppressed matrix element between initial and final states for the RR $\rightarrow$ RL process is constructed by moving through a sequence of off-shell intermediate states, shifted from one another by $\delta \nu \lesssim \delta m$ harmonics.
c) Energy denominators for the virtual states in the process described in b).
Multiplying these energy denominators gives the expression in Eq.~(\ref{eq:A_inter}).
}
\label{fig:HighOrder}
\end{figure}
For simplicity we focus on the case $W \ll \omega \ll \Delta$, where the spread $\delta m$ of the two particle Floquet wave functions in Fourier space is of order $\delta m = O(1)$, see Fig.~\ref{fig:TwoParticleScattering}c,d, and Fig.~\ref{fig:HighOrder}.
The dependence of the two-particle matrix elements on $\nu - \nu'$ is distinguished by the Floquet band indices of the incoming and outgoing states.
In particular, as seen for the special case of the on-shell process depicted in Fig.~\ref{fig:TwoParticleScattering}d, 
for an RR$\rightarrow$ RR (intraband) transition the harmonic-space wave functions give an order-1 contribution (no suppression) to Eq.~(\ref{eq:TimeDomainMatEl}) for $|\nu-\nu'| \lesssim \delta m$.
For an RR$\rightarrow$ RL matrix element, order-1 overlap of the harmonic-space wave functions is achieved for $\frac{\Delta}{\omega} - \delta m \lesssim |\nu-\nu'| \lesssim \frac{\Delta}{\omega} + \delta m$.
Note that this non-suppressed interband scattering matrix element necessarily describes an {\it off-shell} process.
We neglect ``double interband'' processes RR $\rightarrow$ LL, the rates of which are heavily suppressed compared with those of the primary RR $\rightarrow$ RL decay process.


Using the rules above, we see that at high order in perturbation theory it is possible to construct an interband scattering amplitude that avoids the (super) exponential suppression arising from the tiny overlap of harmonic space wave functions that appears in the Born approximation.
At each step in the perturbation theory, $\mathcal{U}$ may connect virtual states with $|\nu-\nu'| \sim \delta m$. 
Therefore, as indicated in Fig.~\ref{fig:HighOrder}b, there is an order $N_{\rm min}\sim \frac{\Delta}{\delta m \omega}$ at which the initial (RR) and final (RL) two-particle states can be connected with no suppression due the harmonic space wave functions.

Viewing the construction of the $N_{\rm min}$-order scattering amplitude as a sequence of steps, each application of $\mathcal{U}$ reduces the harmonic-space separation between the virtual intermediate and the final (on-shell) state.
These successive off-shell intermediates bring energy denominators of larger and larger magnitudes.
After $n$ steps, one particle is transferred from the R to the L band, giving a jump in the sequence of energy denominators.
The terms in this sequence, in units of $\delta m \omega$, are of order $-1,-2,\dots,-n,N_{\mathrm{min}}-n,N_{\mathrm{min}}-n-1,\dots,1$, see Fig.~\ref{fig:HighOrder}c.
We write the corresponding transition amplitude as 
\be
\label{eq:A_inter}A^{(n)}_{\mathrm{inter}} \approx \frac{(-1)^n a_n} {n ! (N_{\mathrm{min}}-n) ! (\delta m \omega )^{N_{\mathrm{min}}}} .
\ee
Here, $a_n \propto U^{N_{\mathrm{min}}}$ contains a sum of matrix elements of the interaction between the intermediate states.
This gives the following rough estimate for the inter-band scattering rate (see Appendix~\ref{sec: estimate rate} for details):
\be
\Gamma_{\mathrm{inter}} \propto  \Big \vert \sum_n A^{(n)}_{\mathrm{inter}} \Big \vert^2 \sim\  \left(\frac{\alpha U}{\Delta} \right)^{\frac{2\Delta}{\delta m \omega}}.
\label{eq:high-order}
\ee
Here, $\alpha$ is an $O(1)$ numerical constant.

Physically, in the regime $W \ll \omega \ll \Delta$ that we have considered, the dominant inter-band scattering processes are associated with $N_{\mathrm{min}}$ scattering events, each one involving the absorption of $\delta m$ energy quanta of $\omega$ from the driving field, such that the total absorbed energy is $\Delta$. We expect such processes to dominate as long as $\omega$ is not too small compared to $W$. 

If $W \gg \omega$, the typical change in $m$ in every scattering event scales with the width of the single-particle Floquet wave functions in harmonic space, $\delta m \sim W/\omega$. 
Hence we would have $N_{\mathrm{min}} \sim \frac{\Delta}{W}$ in Eq.~(\ref{eq:high-order}).
In this regime, we expect both $\Gamma_{\mathrm{inter}}$ and $\Gamma_{\mathrm{intra}}$ to vanish as power laws in $\omega$, since in the static limit the system does not absorb energy from the driving field.
Linear response theory gives that both $\Gamma_{\mathrm{inter}}$ and $\Gamma_{\mathrm{intra}}$ scale as $\omega^2$.
However, $\Gamma_{\mathrm{inter}}$ contains an additional suppression factor, which is exponentially small in $\Delta/W$.
Therefore we expect that the ratio $\Gamma_{\mathrm{inter}}  / \Gamma_{\mathrm{intra}}$ remains small for $\omega\ll W$, as long as $\Delta \gg W,U$.
Hence there is a broad parameter regime where the quasi-steady state persists over many driving periods, even for arbitrarily small driving frequency.

\section{Numerical simulations}
 \label{sec: Numerics}
To further study the many-body dynamics, we have performed exact numerical simulations 
of finite-size systems. 
{To minimize the Hilbert space dimension we consider fermions in these simulations}. We believe that the qualitative behavior does not depend on the quantum statistics, 
but leave the detailed study of the bosonic case to future investigations.

In each simulation, we initialize the system in a Slater determinant state where $N$ momentum states in the right-moving (R) Floquet band are occupied, and the left-moving (L) band is empty. The results do not depend sensitively on which states in the R band are initially occupied, except at very short times.
The particles move on a lattice of $L$ unit cells {(with two sites each)}, with periodic boundary conditions. The largest system we studied contained 8 particles with $L=16$ unit cells.

Figures~\ref{fig:summary}c,d show the {period-averaged} current,
\be
\mathcal{J}(n_T) = \int_{n_T T}^{(n_T+1)T}\!\!\!\! dt'\, \frac{1}{L} \sum_j J'(t') \left\langle  ic^\dagger_{j,B} c^{\vphantom{\dagger}}_{j+1,A} + \textrm{h.c.} \right\rangle_{t'},
\ee
as a function of $n_T$, the number of periods elapsed (see figure caption for model parameters). At time $t=0$, the system carries a current that depends on the initial state. Over a timescale $\tau_{\mathrm{intra}} \sim 10\,T$ the current relaxes to a value $\mathcal{J} \approx \rho/T$, see Fig.~\ref{fig:summary}c (in this simulation, $\rho=0.5$). We ascribe this time scale to relaxation within the R band, while the L band remains almost unpopulated. Examining the occupation numbers of {momentum states} confirms this interpretation (see Appendix~\ref{sec: numerics}); for $t \gtrsim \tau_{\mathrm{intra}}$, the occupation numbers in the R band are approximately uniform and all close to $\rho$.
The average group velocity of states in the R band is $v_g=a/T$. Therefore, the average current $\mathcal{J} = \frac{\rho}{a} v_g$ observed in Fig.~\ref{fig:summary}c is as expected for a uniform particle distribution in the right moving band.

\begin{figure}[t]
\includegraphics[width=0.9 \columnwidth]{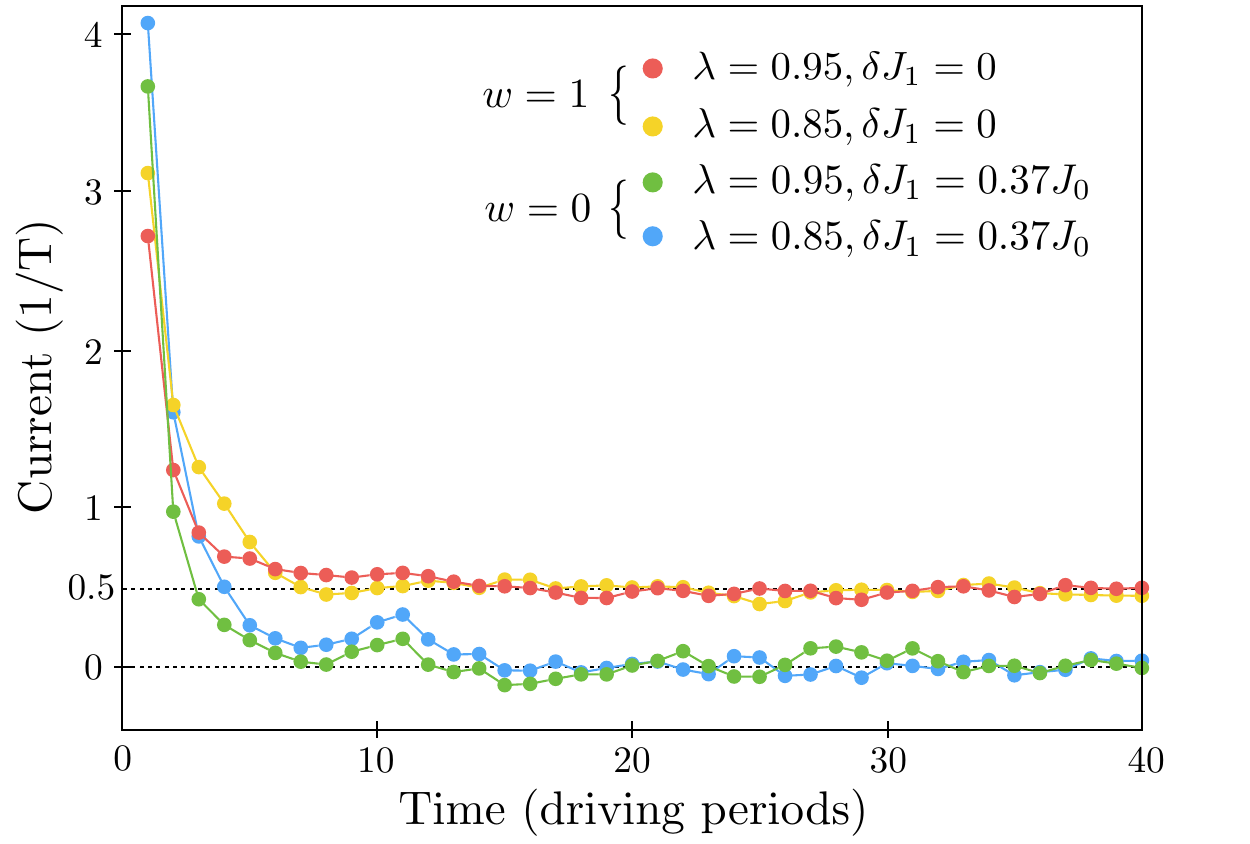}
\caption{Universality of the quasi-steady state.
Time traces of current are shown for four choices of parameter values for the single particle part of the Hamiltonian, Eq.~(\ref{eq: h0}), showcasing the topologically trivial ($w = 0$) and non-trivial ($w = 1$) regimes.
After a transient of several driving periods, the current saturates to the value $\Avg{\mathcal{J}} = w\rho/T$, where $\rho = 0.5$ is the density of particles, per unit cell.
In all cases, the system is initialized by filling all states with crystal momentum values in the range $0 \le k < \pi/a$, in a single Floquet band. The simulation was done with $\omega=0.23J_0$ and $U=2J_0$. }

\label{fig:Robustness}
\end{figure}
To explore the robustness of the universal current in the quasi-steady state, we simulated the dynamics for a variety of driving parameter values.
In Fig.~\ref{fig:Robustness} we show the transient dependence of current as a function of time in both the nontrivial ($w = 1$) and trivial ($w = 0$) regimes.
The data clearly demonstrate that current saturates to a value $\mathcal{J} \approx \rho w/T$ after a few driving periods.

At longer times the current undergoes a much slower decay process toward zero, with a time scale $\tau_{\mathrm{inter}}$.
This decay is visible in the blue trace (corresponding to $\omega = 0.33 J_0$) in Fig.~\ref{fig:summary}d, where $\tau_{\rm inter}$ takes a value on the order a few hundred periods; at lower frequency (red curve, $\omega = 0.13 J_0$), no signs of decay are visible within the simulated time range of 1000 periods.
During decay, the population of the L band gradually increases. {For $t\to\infty$} the occupation numbers of all the momentum states in both bands tend toward $\rho/2$, corresponding to a maximally randomized {infinite-temperature-like} state.

\begin{figure}[t]
\includegraphics[width=1.0 \columnwidth]{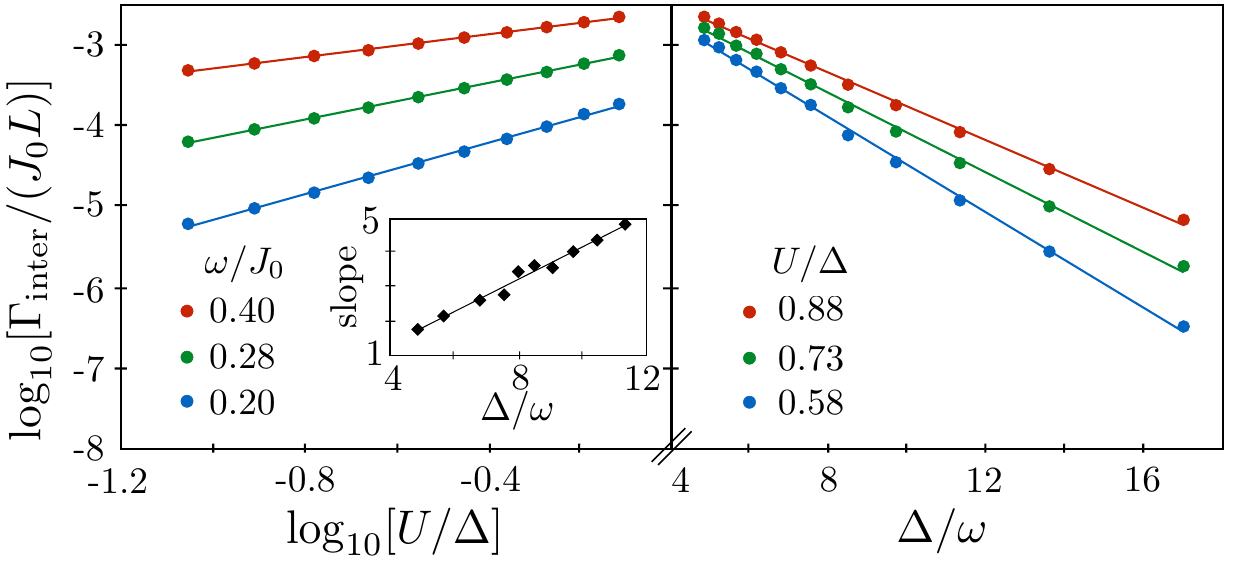}
\caption{Interband excitation rate $\Gamma_{\rm inter}$ obtained from exact numerical evolution of a system with 8 fermions on 32 sites ($L=16$ unit cells).
Left panel: Log-log plot showing dependence of $\Gamma_{\rm inter}/(J_0L)$ on interaction strength $U$, for three fixed values of frequency $\omega$ and for band structure parameter $\lambda = 0.56$. Trend lines are linear fits to the data, confirming a power law dependence consistent with Eq.~(\ref{eq:high-order}).
Inset: Power law exponent as a function of $\omega$, extracted from linear fits to the data.
Right panel: Log-linear plot showing the driving frequency dependence of $\Gamma_{\rm inter}/(J_0L)$,
indicating an exponential dependence on $1/\omega$, for the same model parameters. }
\label{fig:Numerics}
\end{figure}
For times up to several times {$\tau_{\mathrm{intra}}$}, we find that the average (total) occupation number in the L band increases approximately linearly with time. The slope of the linear growth, which we define as $\Gamma_{\mathrm{inter}} \equiv 1/\tau_{\mathrm{inter}}$, is found to be only weakly dependent on system size for $L$ between $10$ and $16$ (see Appendix~\ref{sec: numerics}).

Figure \ref{fig:Numerics} presents the rate $\Gamma_{\mathrm{inter}}$ as a function of model parameters. The dependence of $\Gamma_{\rm inter}$ on $U$ for different values of $\omega$ (with all other parameters fixed) is presented in the left panel of Fig.~\ref{fig:Numerics} in a log-log plot, showing a power-law dependence on $U$. The power depends linearly on $1/\omega$ (left panel, inset). The right panel shows the dependence on $\omega$ at fixed $U$. Clearly, $\Gamma_{\rm inter}$ scales exponentially with $1/\omega$. For the lowest frequencies we studied, corresponding to $\Delta/\omega \approx 16$ (where $\Delta$ is the minimum instantaneous gap), the rate reaches $\sim10^{-7}$ in units of $J_0$, indicating a very long-lived quasi-steady state where only the R band is populated. We also find that the behaviour of $\Gamma_{\rm inter}$ shown in Fig~\ref{fig:Numerics} is not sensitive to the details of the band structure, and persists throughout a wide range of values of the band structure parameter $\lambda$.

Interestingly, the numerically obtained $\Gamma_{\rm inter}(U,\omega)$ is consistent with the form of Eq.~(\ref{eq:high-order}).
This suggests that, within our model, the inter-band relaxation process is dominated by ``multi-photon assisted'' scattering events, where many energy quanta are absorbed from the driving field.


\section{Discussion} 
In this work we studied the dynamics of a periodically-driven many-body system where the driving frequency is much smaller than the instantaneous inter-band gap throughout the driving period. When the system is prepared such that one of the bands is initially empty, while the other is partially occupied, 
a very long intermediate time window emerges in which 
a universal quasi-steady state is realized. The quasi-steady state carries a robust current whose value depends solely on the density of particles and the topological winding number of the underlying single-particle Floquet spectrum.

The combination of periodic driving and interactions leads to rapid heating of the low-energy degrees of freedom of the system, while the large single-particle band gap suppresses (high-energy) interband scattering.
In this way the system showcases a novel paradigm of prethermalization, occuring for low frequency driving, where heating to restricted, highly randomized states allows new topological features to be exposed.

Beyond demonstrating the existence of this interesting new dynamical regime of slowly driven many-body systems, 
our results may be directly applicable to recent experiments on cold atomic fermions~\cite{nakajima2016topological} and bosons~\cite{lohse2015thouless}, where quantized pumping has recently been demonstrated. For the parameter regime stated above, initializing the system with a fractional filling of one of the bands should result in a current carrying quasi-steady state.

The mechanism presented here should apply to systems in higher spatial dimensions, as well.
Generally, the separation of energy scales provides a way for interactions to stabilize non-trivial long lived prethermal states, whose universal properties expose the topological features of a system's underlying Floquet band structure.
For example, recent experiments with cold bosonic atoms in two-dimensional optical lattices 
found that, over an intermediate time window, the Hall conductivity is given by the average band filling times the Chern number of the band in which the particles are initialized~\cite{aidelsburger2015measuring}.
We believe that such behavior can be understood in terms of the mechanism described here, and that our analysis can provide a guide for extending the lifetimes of this interesting prethermal state. 

The same mechanism may apply in adiabatically-driven three dimensional systems, where the Floquet bands can have topological properties beyond those of static bands~\cite{KBRD}. Adding interactions to these systems can lead to novel universal quasi-steady states with quantized response functions, reflecting the topological nature of the Floquet bands. These new states and their response characteristics will be interesting directions to explore in future work.

\begin{acknowledgments}
We thank Dima Abanin, Max Genske, Michael Knap, and Anatoli Polkovnikov for illuminating discussions, and David Cohen for technical support. NL and EB acknowledge financial support from the European Research Council (ERC) under the European Union Horizon 2020 research and innovation programme (grant agreement No 639172). MR gratefully acknowledges the support of the Villum Foundation and the People Programme (Marie Curie Actions) of the European Union, Seventh Framework Programme (FP7/2007-2013) under REA grant agreement PIIF-GA-2013-627838.

\end{acknowledgments}
\appendix

\section{Single particle Floquet wave functions}
\label{sec: single particle}

In this Appendix we elaborate on the extended zone formalism, and
apply it to the single-particle Floquet states. In particular, we
derive the (localized) form of the Floquet wave functions in harmonic space by 
relating the problem to Zener tunneling in a two-band system in a linear potential.

Consider a single-particle Hamiltonian of the form $H_{0}(t)=H_{\mathrm{dc}}+\Lambda e^{i\omega t}+\Lambda^{\dagger}e^{-i\omega t}$.
(The same formalism applies to any time-periodic Hamiltonian.)
We assume that the single particle Hamiltonian is diagonal in momentum space, and focus on the Floquet states $\vert\Psi_{k}(t)\rangle$ for a single value of the crystal momentum $k$.
For simplicity we also assume that the system has only two bands.
The corresponding $2\times 2$  Bloch Hamiltonian is $H_{0,k}(t) = H_{\mathrm{dc},k} + \Lambda^{\vphantom{\dagger}}_{k}e^{i\omega t} + \Lambda^\dagger_{k}e^{-i\omega t}$.

\begin{widetext}
Inserting a Floquet state of the form $\vert\Psi_{k}(t)\rangle=e^{-i\varepsilon_{k}t}\sum_{m}\vert\varphi_{k}^{(m)}\rangle e^{-im\omega t}$
into the time-dependent Schr\"{o}dinger equation, we get that the quasi-energy $\varepsilon_k$ and the Fourier components $\vert \varphi_k^{(m)} \rangle$ satisfy: 

\begin{equation}
\left(\mbox{\mbox{\ensuremath{\begin{array}{ccccc}
\ddots & \Lambda_{k}\\
\Lambda_{k}^{\dagger} & H_{\mathrm{dc},k}+\omega & \Lambda_{k}\\
 & \Lambda_{k}^{\dagger} & H_{\mathrm{dc},k} & \Lambda_{k}\\
 &  & \Lambda_{k}^{\dagger} & H_{\mathrm{dc},k}-\omega & \Lambda_{k}\\
 &  &  & \Lambda_{k}^{\dagger} & \ddots
\end{array}}}}\right)\mbox{\ensuremath{\left(\begin{array}{c}
\vdots\\
\vert\varphi_{k}^{(-1)}\rangle\\
\vert\varphi_{k}^{(0)}\rangle\\
\vert\varphi_{k}^{(1)}\rangle\\
\vdots
\end{array}\right)}\ =\ \ensuremath{\varepsilon_{k}\left(\begin{array}{c}
 \vdots\\
 \vert\varphi_{k}^{(-1)}\rangle\\
 \vert\varphi_{k}^{(0)}\rangle\\
 \vert\varphi_{k}^{(1)}\rangle\\
 \vdots
\end{array}\right)}.}\label{eq:extH}
\end{equation}
\end{widetext}
The operator on the left
hand side of this equation is the Floquet ``extended zone'' operator, or ``extended Hamiltonian,'' $\mathcal{H}_{0,k}$.
Note that $\mathcal{H}_{0.k}$ has a block-tridiagonal form, analogous to that of a nearest-neighbor tight binding Hamiltonian.
In this picture, the terms proportional to $\omega$ on the diagonal correspond to a linear potential on the $m$-lattice.

In the adiabatic limit, $\omega\ll\Delta$, we first solve the problem with $\omega=0$.
The  diagonal ``linear potential'' is later introduced as a perturbation.
For $\omega=0$, the problem is translationally invariant in harmonic
($m$) space; we can solve it by Fourier transforming along the $m$-direction.
This amounts to transforming from frequency back to time domain (where $t$ plays the role of ``momentum'' for the $m$-space tight-binding problem).
The secular equation then becomes
\begin{equation}
H_{0}(t)\vert\Psi_{\alpha}(t)\rangle=E_{\alpha}(t)\vert\Psi_{\alpha}(t)\rangle,
\end{equation}
which is nothing but the Schr\"{o}dinger equation for the \emph{instantaneous}
eigenstates and eigenenergies. Here, $\alpha=\mbox{R},\mbox{L}$ is
the band index. The two bands are separated by a gap, $|E_{L,k}(t)-E_{R,k}(t)|\ge\Delta$.

\begin{figure}[t]
\includegraphics[width=1.0 \columnwidth]{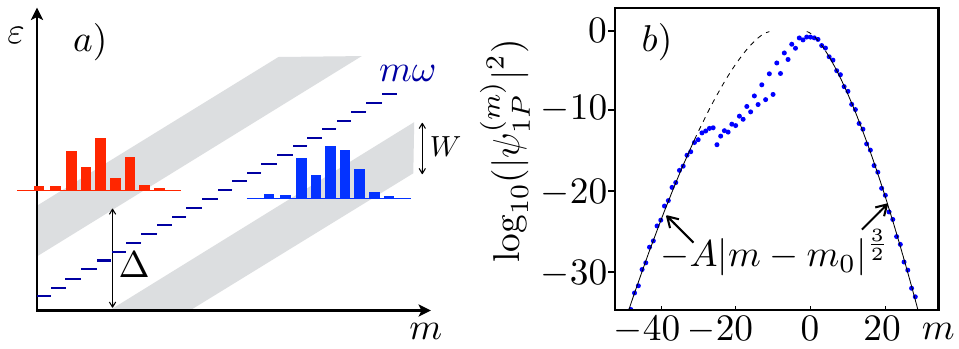}
\caption{ Effective Zener tunneling problem for the Floquet states in harmonic space. a)  The Floquet problem, Eq.~(\ref{eq:extH}), is equivalent to a tight-binding problem in harmonic ($m$) space, with a linear potential term $m \omega$. The effective bands have a characteristic width $W$ and gap $\Delta$. The gray regions indicate the ``classically allowed'' regions for each fixed value of ``energy,'' 
as determined by the sum of the band energies and the linear potential. The Floquet wave functions in the two bands (illustrated by the red and blue bars) are strongly localized in the classically allowed regions, and decay rapidly beyond these regions. b) Modulus squared of a representative Floquet wave function as a function of the Fourier index $m$. Sufficiently far from the maximum, the wave function decays as $\exp(-A|m-m_{0}|^{3/2})$, where $A$ and $m_0$ are constants. }
\label{fig:zener}
\end{figure}

Next, we consider the effect of the linear potential term. 
The situation is illustrated in Fig.~\ref{fig:zener}a. The problem is equivalent to the well-known Zener tunneling problem of a two-band semiconductor in an
electric field. The states in both bands become localized in the ``classically
allowed region'' in harmonic space, whose characteristic size is of
order $W_{\alpha}/\omega$ (where $W_{\alpha}=\mbox{max}_{k,t}\{E_{\alpha k}(t)\}-\mbox{min}_{k,t}\{E_{\alpha k}(t)\}$ is an effective bandwidth).

Within this picture, states in the R and L bands that are close in quasienergy (within $\sim\omega$
of each other) are separated in harmonic space by a spacing of the
order of $\Delta/\omega$. The tails of the wave functions decay rapidly
into the classically forbidden region; in the limit $\omega\ll W_{\alpha}$,
there is a broad region where the wave functions decay as $e^{-A|m-m_{0}|^{3/2}}$ (see Fig.~\ref{fig:zener}b),
where $A$ is a band structure dependent dimensionless constant and
$m_{0}$ is the boundary of the classically forbidden region. (This
form is expected from a Wentzel\textendash Kramers\textendash Brillouin
treatment of the problem, where $m$ is treated as a continuous variable). At asymptotically long distances, larger than several times $W_{\alpha}/\omega$,
the form of the wave function crosses over to $e^{-B|m-m_{0}|\log|m-m_{0}|}$,
where $B$ is another dimensionless constant.

The tunneling matrix element between the two bands is hence strongly
suppressed in the adiabatic limit. The hybridization between the two
bands is expected to be very small, unless their energies are very
close to each other. As $k$ varies, the bands nearly cross at a set
of $k$ points (see Fig.~1b in the main text); at these crossing points
the two bands hybridize, and there is an avoided crossing gap whose
magnitude is exponentially small in the adiabatic limit. For a generic
band structure, the hybridization between the bands is significant only within exponentially
small regions in $k$ space around the near-crossing points. These hybridizations are responsible for the bright lines appearing in Fig.~2b in the main text.

\section{Stroboscopic dynamics using the extended zone Hamiltonian}
\label{sec: stroboscopic}
The extended zone Hamiltonian $\mathcal{H}$ in harmonic space (shown
in Eq.~(\ref{eq:extH}) for the single-particle case) is designed
such that its spectrum and eigenstates correspond to the quasi-energies
and Floquet states, respectively. It can also be used to generate the
\emph{stroboscopic dynamics} at times $t=nT$, where $n$ is an integer,
starting from an arbitrary initial state.

In order to see this, consider a Floquet state solution of the time-dependent
Schr\"{o}dinger equation generated by the physical Hamiltonian $H(t)$,
of the form $\vert\psi(t)\rangle=e^{-i\varepsilon t}\sum_{m}\vert\varphi^{(m)}\rangle e^{-im\omega t}$.
Compare this state to a solution of the auxiliary-time Schr\"{o}dinger equation in harmonic (extended) space, 
generated by $\mathcal{H}$ (see main text):
\begin{equation}
\varphi_{\mathrm{ext}}(\tau)=e^{-i\varepsilon \tau}(\dots,\vert\varphi^{(-1)}\rangle,\vert\varphi^{(0)}\rangle,\vert\varphi^{(1)}\rangle,\dots)^T.
\end{equation}
 (We use Dirac bra-ket notation for states in the physical Hilbert
space, whereas vectors in the extended space are written without
Dirac notation.) We relate the 
``evolved'' state in the extended space to a state
in the physical Hilbert space at stroboscopic times $t = \tau = NT$ by summing over the harmonic components
of $\varphi_{\mathrm{ext}}(\tau)$: $\vert\varphi_{\mathrm{ext}}(t = NT)\rangle=e^{-i\varepsilon NT}\sum_{m}\vert\varphi^{(m)}\rangle$.
At these stroboscopic times, 
we find that $\vert\psi(t=nT)\rangle$ and $\vert\varphi_{\mathrm{ext}}(t=nT)\rangle$
coincide. 

The reasoning above can be extended to the time evolution of an arbitrary
initial state, $\vert\psi_{0}\rangle$. This is done by expanding
$\vert\psi_{0}\rangle$ in terms of Floquet eigenstates, considering
the time evolution with respect to either $H(t)$ or $\mathcal{H}$,
and comparing the time-evolved wave functions at times $t=nT$.

\section{Construction of non-interacting (``free'') many-body Floquet states}\label{appendix:FreeStates}

We build up the non-interacting $N$-particle state $\phi_{\bm{\xi}}$ one harmonic at a time.
First, we define a single particle Floquet state $\phi_{{\rm 1P},\xi_i}$, as in Eq.~(\ref{eq:phi_vector}), for each $\xi_i = \{k_i,\alpha_i\}$, with $i = 1 \ldots N$.
As used throughout the main text, the single particle quasienergies $\epsilon_{{\rm 1P},\xi_i}$ are taken in the interval  $0\leq \epsilon_{{\rm 1P},\xi_i} < 2\pi/T$.
Within this convention we define a set of creation operators $\hat\straightphi_{\xi}^{(m')\dagger}$, for all $m'$, using the identification $|\varphi^{(m')}_{{\rm 1P}, \xi}\rangle=\hat\straightphi^{(m') \dagger}_{\xi}|0\rangle $.
Our analysis holds for both bosons and fermions, where the creation and annihilation operators respectively satisfy the commutation (+) and anticommutation ($-$) relations $[\hat{\straightphi}_{\xi}^{(m)}, \hat{\straightphi}_{\xi'}^{(m')}]_\pm = [\hat{\straightphi}_{\xi}^{(m)\dagger}, \hat{\straightphi}_{\xi'}^{(m')\dagger}]_\pm = 0$ and $[\hat{\straightphi}_{\xi}^{(m)}, \hat{\straightphi}_{\xi'}^{(m')\dagger}]_\pm = \Amp{\varphi^{(m)}_{{\rm 1P}, \xi}}{\varphi^{(m')}_{{\rm 1P}, \xi'}}$.

The $m$-th Fourier component of the many-body state $\phi_{\bm{\xi}}$ is determined by a convolution over the single particle harmonics:
\begin{equation}
|\phi^{(m)}_{\bm{\xi}}\rangle= \sum_{\left\{m_i\right\}} \delta_{m,\, \sum_i\! m_i} \left[\prod_{i=1}^N \hat{\straightphi}^{(m_i) \dagger}_{\xi_i}\right]|0\rangle.
\end{equation}
In the above, the sum extends over all sets of N integers $m_i$, for $i=1,..,N$.
Note that the quasi-energy of the many-body state $\phi_{\bm{\xi}}$ is $\epsilon_{\bm{\xi}}=\sum_{i=1}^N\epsilon_{{\rm 1P}, \xi_i}$, with the convention $0\leq \epsilon_{{\rm 1P},\xi_i} < 2\pi/T$ specified above.
Thus  $\epsilon_{\bm{\xi}}$ is uniquely specified, and generically falls outside the fundamental Floquet-Brillouin zone.  

\section{Estimate of the interband scattering rate}
\label{sec: estimate rate}

Here we describe the steps leading to the estimate of the interband scattering rate [Eq.~(6) of the main text]. We begin with the term in the expression for the amplitude of order $N_{\mathrm{min}}$, in which a particle is scattered from the L to the R band after $n$ steps, with  $0\le n \le N_{\mathrm{min}} - 1$. This amplitude (in the limit $W \ll \omega$) is given by
\be
A^{(n)}_{\mathrm{inter}} \approx \frac{(-1)^n a_n} {n ! (N_{\mathrm{min}}-n) ! (\delta m \omega )^{N_{\mathrm{min}}}}.
\ee
Using Stirling's formula for $n,N_{\mathrm{min}}\gg 1$, we may replace $n ! (N_{\mathrm{min}}-n) ! \approx e^{N_{\mathrm{min}} [ \log (N_{\mathrm{min}}) - f(n/N_{\mathrm{min}})]}$, where $f(x) = 1+x\log(x) + (1-x)\log(1-x)$. The function $f(x)$ is bounded and $f(x)=O(1)$ for $0 \le x \le 1$.
Therefore we approximate the factor $e^{-N_{\rm min}f(n/N_{\rm min})}$ by $\beta_1^{-N_{\rm min}}$
to leading exponential accuracy in the limit of large $N_{\mathrm{min}}$, where $\beta_1 = O(1)$ is some constant.

In order to get a rough estimate of the interband scattering rate $\Gamma_{\mathrm{inter}}$, we must examine the factor $a_n$ in Eq.~(6), which contains a sequence of matrix elements of the interaction term $\mathcal{U}$ between the initial, intermediate, and final states. The dominant dependence of $a_n$ on $N_{\mathrm{min}}$ is expected to be of the form $a_n \sim (\beta_2 U) ^{N_{\mathrm{min}}}$, where $\beta_2 = O(1)$. Hence, summing the amplitudes $A^{(n)}_{\mathrm{inter}}$ over $n$, taking the modulus of the square, and using $N_{\mathrm{min}}\sim \frac{\Delta}{\delta m \omega}$, we get to the estimate for $\Gamma_{\mathrm{inter}}$ quoted in Eq.~(6) of the main text. While this estimate is rather rough, our numerical results for $\Gamma_{\rm inter}$ show excellent agreement with the form predicted in Eq.~(6).

\section{Single-particle resonances}
\label{sec:resonance}

\begin{figure}[t]
\includegraphics[width=1.05\columnwidth]{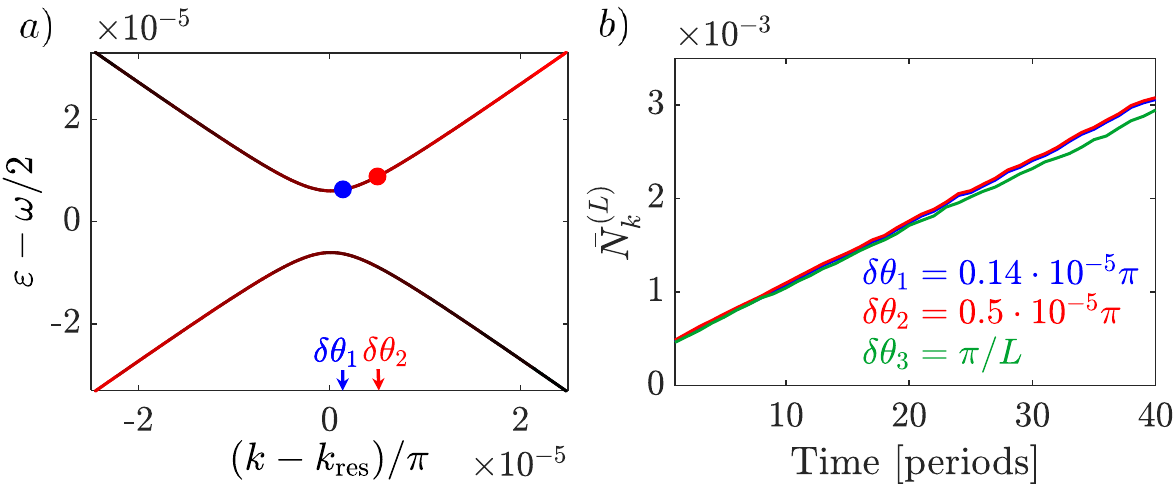}
\caption{a) Single-particle Floquet spectrum in the vicinity of an avoided crossing between the two counter-propagating bands. In this calculation, $J_0=1.5$, $J_1 = 1$, $V_1 = 3$, $U = 3$, and $\omega = 0.3$. The single-particle resonance is at $k_{\mathrm{res}} = 1.2856033\pi$. We use twisted boundary conditions to tune one of the allowed $k$ points in our $L=16$ system near the resonance; the $k$ points nearest to $k_{\mathrm{res}}$, corresponding to twist angles $\delta \theta_1$ and $\delta \theta_2$, are shown by blue and red dots, respectively. b) Total occupation number of the left moving Floquet band, excluding the $k$ point nearest to $k_{\mathrm{res}}$, as a function of time, for three different twist angles, $\delta \theta_{1,2,3}$. Two of the twist angles, $\delta\theta_1$ and $\delta\theta_2$, are such that there is a $k$ point close to $k_{\mathrm{res}}$ (blue and red curves, respectively), while for the third angle, $\delta\theta_3$, none of the $k$ points are near $k_{\mathrm{res}}$ (green curve). This is a simulation of $N=8$ particles in an $L=16$ system. The interaction strength is $U=3$. }
\label{fig: resonance}
\end{figure}

Our many-body simulations reveal that the interband scattering rate is well-described by Eq.~(\ref{eq:high-order}). In particular, for a fixed interaction strength, the rate depends exponentially on $\Delta/\omega$. Since the single-particle avoided crossing gap that forms at a crossing point of the counter-propagating Floquet bands is similarly exponentially small in $\Delta/\omega$, one may wonder whether the interband scattering rate is dominated by many-body processes of the type described in Sec.~\ref{sec: estimate rate}, or by the single-particle resonances.
 The latter have a pronounced effect on the two-body inter-band scattering rate within the Born approximation; see Fig.~\ref{fig:TwoParticleScattering}.


In this Appendix, we test the effect of the single-particle resonances on the many-body dynamics. In our finite size many-body simulation, momentum is quantized in units of $2\pi/L$, with $L=10$ to $16$.
Generically, none of the allowed momentum points are close to the single-particle resonances, which affect only a small region in momentum space.
To assess the effect of the resonances, we performed simulations with twisted boundary conditions, such that one of the allowed $k$ points is tuned to be close to a single-particle resonance.
Such tuning should greatly over-estimate the effect of the single-particle resonances in our finite system, compared to the thermodynamic limit:
in the finite system, a fraction $1/L$ of the momentum points are strongly affected by the proximity to a resonance point, whereas in the thermodynamic limit only 
an exponentially small fraction $\Delta_{\mathrm{res}}/2\pi v$ of the Brillouin zone is significantly affected by the resonances.
Here $\Delta_{\mathrm{res}}$ is the (exponentially small) single-particle avoided crossing gap due to a multi-photon resonance, and $v$ the group velocity of the bands near the crossing point.

Fig.~\ref{fig: resonance} shows results of simulations with three different twist angles, $\theta_i$, $i=1,2,3$, such that the allowed single-particle momenta on the ring are $k_n = 2\pi n/L + \theta_i$ ($n$ is an integer). We parametrize $\theta_i = \theta_0 + \delta \theta_i$, where $\theta_0$ is the twist angle necessary for one of the $k$ points to hit one of the resonances, and $\delta\theta_{1,2,3} = \{0.14\cdot 10^{-5}, 0.5\cdot 10^{-5}, 1/L\}\times \pi$, respectively, with $L=16$.

Fig.~\ref{fig: resonance}a presents the single-particle Floquet spectrum zoomed in to the vicinity of one of the resonance points. The position of the nearest allowed $k$ point is shown for $\delta\theta_{1,2}$; the twist angles have been tuned to one part in $10^6$ in order to place this $k$ point close to the resonance. In contrast, $\delta\theta_3$ is such that the all the $k$ points are as far as possible from the resonance. As before, we initialize the system with $N=L/2=8$ particles in the $\mathrm{R}$ band and evolve it in time. For $\delta\theta_{1,2}$, the initial conditions are chosen such that the $k$ point near the resonance is initially unoccupied~\footnote{This is important since, at this point, the Floquet eigenstates are mixtures of states from the upper and lower instantaneous bands; we would like to initialize the system with states that are mostly from the instantaneous lower energy band.}.

Fig.~\ref{fig: resonance}b shows the total occupation of the $\mathrm{L}$ band as a function of time (excluding the $k_{n=0}$ point, which for $\delta\theta_{1,2}$ is near the resonance) for the three values of the twist angle $\delta\theta_i$. The results for $\delta\theta_{1}$ and $\theta_{2}$ are essentially identical, despite the fact that for $\delta\theta_1$, one of the $k$ points is about three times closer to the resonance compared to the nearest point for $\delta\theta_{2}$. The average inter-band scattering rate over $40$ periods for $\delta\theta_3$, where there is no $k$ point close to any of the resonances, is within about $3\%$ the rate for $\delta\theta_{1,2}$. We further examined the current carried by the quasi-steady state that forms after a few periods for the three boundary conditions (not shown), and found that the average current from $20$ to $40$ periods is the same for the different $\delta\theta_i$'s to within about $2\%$, and their root-mean-square temporal fluctuations around the average value are similar, as well.

\begin{figure}[t]
\includegraphics[width=8.8cm]{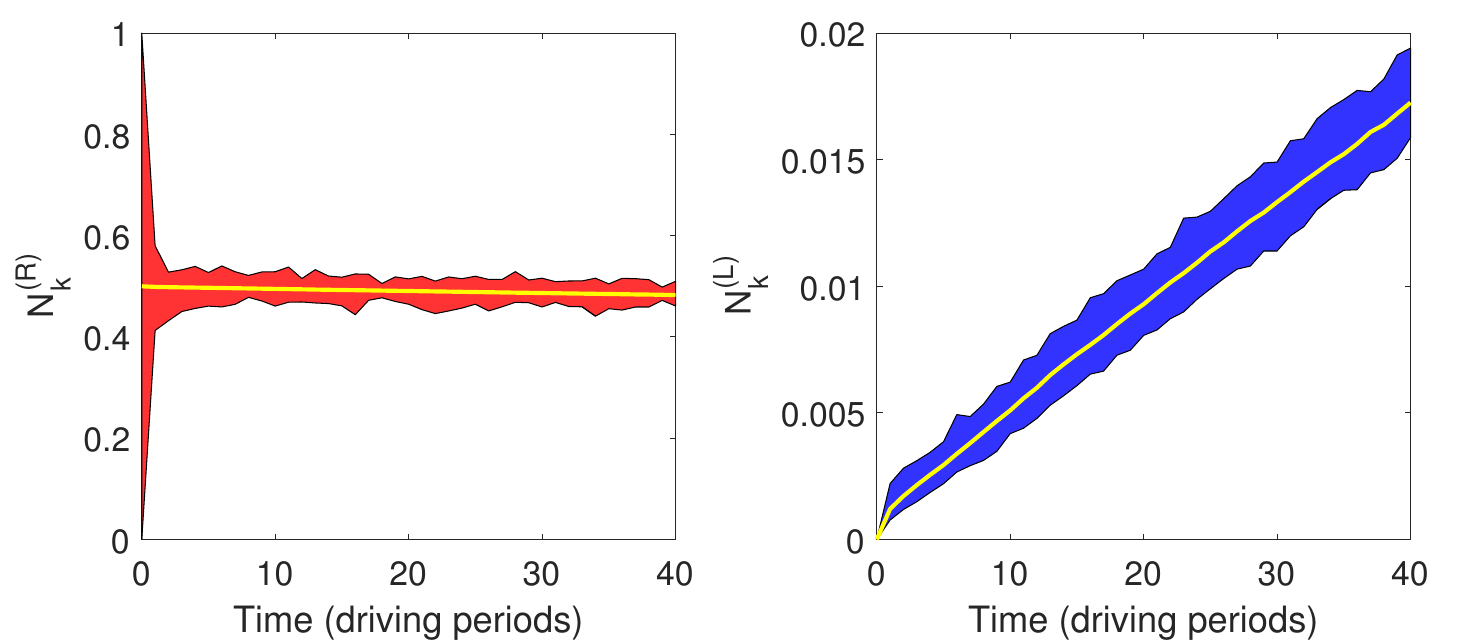}
\caption{Particle distribution in single particle Floquet states in the right and left moving Floquet bands, $N_k^{(R)}$ and $N_k^{(L)}$, as a function of time.  In both figures, we indicate only the maximal and minimal occupations, $\max_k N_k^{(\alpha)}$ and $\min_k N_k^{(\alpha)}$, by shading the area between them. In both plots the yellow line indicates the average occupation $\frac{1}{L}\sum_k N_k^{(\alpha)}$ within the given band. As can be seen from the left figure, the distribution in the right moving band relaxes to an approximately uniform distribution within a short time scale of a few driving periods. The right figure clearly shows the interband scattering rate from the right to the left moving Floquet band. The parameters for this figure were $\omega=0.27 J_0$, $\lambda=0.56$, $U=1.67 J_0$, $L=16$, and $\rho=0.5$.}
\label{fig: occupations}
\end{figure}

These results suggest that the single-particle resonances have little effect on the inter-band scattering rate or the quasi-steady state that forms at intermediate times. As a further evidence for this, note that the polynomial dependence of the rate on the interaction strength, $U$, in Eq.~(\ref{eq:high-order}) suggests that the dominant inter-band scattering is a high-order process in the interaction, rather than a single-particle process.

Why do the single-particle resonances affect the many-body dynamics so weakly?
We propose the following physical picture.
When a particle in the lower instantaneous band is scattered into a momentum state near a resonance, a time of order $1/\Delta_{\mathrm{res}}$ is needed for a transition to the upper band to occur.
This time is exponentially long in $\Delta/\omega$, and is therefore typically much longer than the time for intra-band scattering [which scales as $1/U^2$, see Eq.~(\ref{eq: golden})].
Therefore it is likely that the particle will be scattered to other states within the lower band before it has time to complete the transition.
In such a scattering event, phase coherence between the states of the two bands at $k_{\mathrm{res}}$ 
is destroyed, inhibiting the transition to the higher band.

\section{Further details of the numerical simulations}
\label{sec: numerics}

\begin{figure}[t]
\includegraphics[width=7cm]{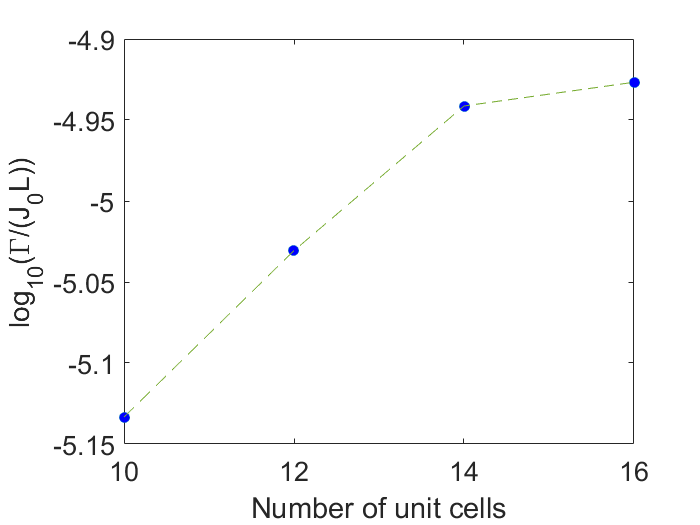}
\caption{Size dependence of the interband scattering rate normalized by the size of the system, $\Gamma/(J_0L)$. 
As the number of unit cells, $L$, is changed, the density of particles is held fixed at $\rho=0.5$.
The parameters used for this figure were $\omega=0.3$, $\lambda=0.56$, $U=2$.  }
\label{fig: finite_size}
\end{figure}

In our numerical simulations, the many-body wave function $|\Psi(t)\rangle$ is time-evolved numerically, using the Hamiltonian $H(t)$ [Eqs.~(1,2)], for up to 1000 driving periods. The simulations are performed using a finite time step, $\Delta t$, and using a Trotter-Suzuki decomposition of the evolution operator within each time step.  Most of the simulations were done with $\Delta t = T/500$.  For $\omega<0.2J_0$, we used $\Delta t = T/850$. We have verified the results do not change upon decreasing $\Delta t$ further, even for the longest times simulated.

To extract the inter-band rates plotted in Fig.~3 of the main text, we computed the distribution of particles in the different single-particle Floquet states at times which correspond to integer multiples of the driving period. We thus calculate $N_k^{(\alpha)}(t=mT)=\langle\Psi(mT)|\hat{\psi}_{k,\alpha}^\dagger\hat{\psi}_{k,\alpha} | \Psi(mT)\rangle$ where $\hat{\psi}^\dagger_{k,\alpha}$ is the creation operator for the Floquet state $|\psi(0)_{k,\alpha}\rangle$ with momentum $k$, and the index $\alpha=$R, L indicates the right or left moving Floquet band.

Typical particle distributions in the Floquet bands are plotted in Fig.~\ref{fig: occupations}.
States within the right moving band quickly become nearly equally populated, with probabilities close to 0.5 (corresponding to the density $\rho=N/L=0.5$, taken in the simulation).
This indicates the establishment of a quasi-infinite-temperature state, restricted to the right moving (R) band.
After an initial transient of a few driving periods, the population in the left moving (L) band increases linearly with time, with a small rate,  while the population in the right moving band decreases with the same rate. The rates shown in Fig.~3 of the main text were obtained by considering the rate of increase of the average population in the left moving band (slope of the yellow line in the right panel of Fig.~\ref{fig: occupations}), $\Gamma_{\rm inter}=\frac{1}{L}\sum_k \left[N_k^{(L)}(t=mT) - N_k^{(L)}(t=m_0T)\right]/\left[(m-m_0)T\right]$ with $m=20$ and $m_0=5$.

The largest system that we could reach with our numerical simulations included 8 particles on 32 sites (i.e., $L=16$ unit cells and density $\rho=0.5$). To verify that the rates reported in Fig.~3 of the main text do not suffer from substantial finite size effects, we studied the size dependence of the interband scattering rate, normalized to the length
of the system,  $\Gamma_{\rm inter}/(J_0L)$, while keeping the density fixed. For the parameter regimes plotted in Fig.~3 of the main text, we found that the size-normalized rate is only weakly size dependent between $L=10$ and $L=16$, and appears to be saturating by $L=16$. This indicates that at $L=16$ finite size effects are small and do not change the results qualitatively. A representative plot of the finite size dependence of the size-normalized rate can be found in Fig.~\ref{fig: finite_size}.

\bibliography{floquet_refs}

\begin{thebibliography}{72}%
\makeatletter
\providecommand \@ifxundefined [1]{%
 \@ifx{#1\undefined}
}%
\providecommand \@ifnum [1]{%
 \ifnum #1\expandafter \@firstoftwo
 \else \expandafter \@secondoftwo
 \fi
}%
\providecommand \@ifx [1]{%
 \ifx #1\expandafter \@firstoftwo
 \else \expandafter \@secondoftwo
 \fi
}%
\providecommand \natexlab [1]{#1}%
\providecommand \enquote  [1]{``#1''}%
\providecommand \bibnamefont  [1]{#1}%
\providecommand \bibfnamefont [1]{#1}%
\providecommand \citenamefont [1]{#1}%
\providecommand \href@noop [0]{\@secondoftwo}%
\providecommand \href [0]{\begingroup \@sanitize@url \@href}%
\providecommand \@href[1]{\@@startlink{#1}\@@href}%
\providecommand \@@href[1]{\endgroup#1\@@endlink}%
\providecommand \@sanitize@url [0]{\catcode `\\12\catcode `\$12\catcode
  `\&12\catcode `\#12\catcode `\^12\catcode `\_12\catcode `\%12\relax}%
\providecommand \@@startlink[1]{}%
\providecommand \@@endlink[0]{}%
\providecommand \url  [0]{\begingroup\@sanitize@url \@url }%
\providecommand \@url [1]{\endgroup\@href {#1}{\urlprefix }}%
\providecommand \urlprefix  [0]{URL }%
\providecommand \Eprint [0]{\href }%
\providecommand \doibase [0]{http://dx.doi.org/}%
\providecommand \selectlanguage [0]{\@gobble}%
\providecommand \bibinfo  [0]{\@secondoftwo}%
\providecommand \bibfield  [0]{\@secondoftwo}%
\providecommand \translation [1]{[#1]}%
\providecommand \BibitemOpen [0]{}%
\providecommand \bibitemStop [0]{}%
\providecommand \bibitemNoStop [0]{.\EOS\space}%
\providecommand \EOS [0]{\spacefactor3000\relax}%
\providecommand \BibitemShut  [1]{\csname bibitem#1\endcsname}%
\let\auto@bib@innerbib\@empty
\bibitem [{\citenamefont {Klitzing}\ \emph {et~al.}(1980)\citenamefont
  {Klitzing}, \citenamefont {Dorda},\ and\ \citenamefont
  {Pepper}}]{vonKlitzing1980}%
  \BibitemOpen
  \bibfield  {author} {\bibinfo {author} {\bibfnamefont {K.~v.}\ \bibnamefont
  {Klitzing}}, \bibinfo {author} {\bibfnamefont {G.}~\bibnamefont {Dorda}}, \
  and\ \bibinfo {author} {\bibfnamefont {M.}~\bibnamefont {Pepper}},\
  }\bibfield  {title} {\enquote {\bibinfo {title} {New method for high-accuracy
  determination of the fine-structure constant based on quantized hall
  resistance},}\ }\href@noop {} {\bibfield  {journal} {\bibinfo  {journal}
  {Phys. Rev. Lett.}\ }\textbf {\bibinfo {volume} {45}},\ \bibinfo {pages}
  {494--497} (\bibinfo {year} {1980})}\BibitemShut {NoStop}%
\bibitem [{\citenamefont {Laughlin}(1981)}]{Laughlin1981}%
  \BibitemOpen
  \bibfield  {author} {\bibinfo {author} {\bibfnamefont {R.~B.}\ \bibnamefont
  {Laughlin}},\ }\bibfield  {title} {\enquote {\bibinfo {title} {Quantized hall
  conductivity in two dimensions},}\ }\href@noop {} {\bibfield  {journal}
  {\bibinfo  {journal} {Phys. Rev. B}\ }\textbf {\bibinfo {volume} {23}},\
  \bibinfo {pages} {5632--5633} (\bibinfo {year} {1981})}\BibitemShut {NoStop}%
\bibitem [{\citenamefont {Oka}\ and\ \citenamefont {Aoki}(2009)}]{Oka2009}%
  \BibitemOpen
  \bibfield  {author} {\bibinfo {author} {\bibfnamefont {Takashi}\ \bibnamefont
  {Oka}}\ and\ \bibinfo {author} {\bibfnamefont {Hideo}\ \bibnamefont {Aoki}},\
  }\bibfield  {title} {\enquote {\bibinfo {title} {Photovoltaic hall effect in
  graphene},}\ }\href {\doibase 10.1103/PhysRevB.79.081406} {\bibfield
  {journal} {\bibinfo  {journal} {Phys. Rev. B}\ }\textbf {\bibinfo {volume}
  {79}},\ \bibinfo {pages} {081406} (\bibinfo {year} {2009})}\BibitemShut
  {NoStop}%
\bibitem [{\citenamefont {Inoue}\ and\ \citenamefont
  {Tanaka}(2010)}]{Inoue2010}%
  \BibitemOpen
  \bibfield  {author} {\bibinfo {author} {\bibfnamefont {Jun-ichi}\
  \bibnamefont {Inoue}}\ and\ \bibinfo {author} {\bibfnamefont {Akihiro}\
  \bibnamefont {Tanaka}},\ }\bibfield  {title} {\enquote {\bibinfo {title}
  {Photoinduced transition between conventional and topological insulators in
  two-dimensional electronic systems},}\ }\href@noop {} {\bibfield  {journal}
  {\bibinfo  {journal} {Phys. Rev. Lett.}\ }\textbf {\bibinfo {volume} {105}},\
  \bibinfo {pages} {017401} (\bibinfo {year} {2010})}\BibitemShut {NoStop}%
\bibitem [{\citenamefont {Kitagawa}\ \emph {et~al.}(2010)\citenamefont
  {Kitagawa}, \citenamefont {Berg}, \citenamefont {Rudner},\ and\ \citenamefont
  {Demler}}]{KBRD}%
  \BibitemOpen
  \bibfield  {author} {\bibinfo {author} {\bibfnamefont {Takuya}\ \bibnamefont
  {Kitagawa}}, \bibinfo {author} {\bibfnamefont {Erez}\ \bibnamefont {Berg}},
  \bibinfo {author} {\bibfnamefont {Mark}\ \bibnamefont {Rudner}}, \ and\
  \bibinfo {author} {\bibfnamefont {Eugene}\ \bibnamefont {Demler}},\
  }\bibfield  {title} {\enquote {\bibinfo {title} {Topological characterization
  of periodically driven quantum systems},}\ }\href@noop {} {\bibfield
  {journal} {\bibinfo  {journal} {Phys. Rev. B}\ }\textbf {\bibinfo {volume}
  {82}},\ \bibinfo {pages} {235114} (\bibinfo {year} {2010})}\BibitemShut
  {NoStop}%
\bibitem [{\citenamefont {Lindner}\ \emph {et~al.}(2011)\citenamefont
  {Lindner}, \citenamefont {Refael},\ and\ \citenamefont
  {Galitski}}]{Lindner2011}%
  \BibitemOpen
  \bibfield  {author} {\bibinfo {author} {\bibfnamefont {N.~H.}\ \bibnamefont
  {Lindner}}, \bibinfo {author} {\bibfnamefont {G.}~\bibnamefont {Refael}}, \
  and\ \bibinfo {author} {\bibfnamefont {V.}~\bibnamefont {Galitski}},\
  }\bibfield  {title} {\enquote {\bibinfo {title} {Floquet topological
  insulator in semiconductor quantum wells},}\ }\href {\doibase
  10.1038/nphys1926} {\bibfield  {journal} {\bibinfo  {journal} {Nat. Phys.}\
  }\textbf {\bibinfo {volume} {7}},\ \bibinfo {pages} {490--495} (\bibinfo
  {year} {2011})}\BibitemShut {NoStop}%
\bibitem [{\citenamefont {Lindner}\ \emph {et~al.}(2013)\citenamefont
  {Lindner}, \citenamefont {Bergman}, \citenamefont {Refael},\ and\
  \citenamefont {Galitski}}]{Lindner2013}%
  \BibitemOpen
  \bibfield  {author} {\bibinfo {author} {\bibfnamefont {Netanel~H.}\
  \bibnamefont {Lindner}}, \bibinfo {author} {\bibfnamefont {Doron~L.}\
  \bibnamefont {Bergman}}, \bibinfo {author} {\bibfnamefont {Gil}\ \bibnamefont
  {Refael}}, \ and\ \bibinfo {author} {\bibfnamefont {Victor}\ \bibnamefont
  {Galitski}},\ }\bibfield  {title} {\enquote {\bibinfo {title} {Topological
  floquet spectrum in three dimensions via a two-photon resonance},}\ }\href
  {\doibase 10.1103/PhysRevB.87.235131} {\bibfield  {journal} {\bibinfo
  {journal} {Phys. Rev. B}\ }\textbf {\bibinfo {volume} {87}},\ \bibinfo
  {pages} {235131} (\bibinfo {year} {2013})}\BibitemShut {NoStop}%
\bibitem [{\citenamefont {Gu}\ \emph {et~al.}(2011)\citenamefont {Gu},
  \citenamefont {Fertig}, \citenamefont {Arovas},\ and\ \citenamefont
  {Auerbach}}]{Gu11}%
  \BibitemOpen
  \bibfield  {author} {\bibinfo {author} {\bibfnamefont {Zhenghao}\
  \bibnamefont {Gu}}, \bibinfo {author} {\bibfnamefont {H.~A.}\ \bibnamefont
  {Fertig}}, \bibinfo {author} {\bibfnamefont {Daniel~P.}\ \bibnamefont
  {Arovas}}, \ and\ \bibinfo {author} {\bibfnamefont {Assa}\ \bibnamefont
  {Auerbach}},\ }\bibfield  {title} {\enquote {\bibinfo {title} {Floquet
  spectrum and transport through an irradiated graphene ribbon},}\ }\href
  {\doibase 10.1103/PhysRevLett.107.216601} {\bibfield  {journal} {\bibinfo
  {journal} {Phys. Rev. Lett.}\ }\textbf {\bibinfo {volume} {107}},\ \bibinfo
  {pages} {216601} (\bibinfo {year} {2011})}\BibitemShut {NoStop}%
\bibitem [{\citenamefont {Kitagawa}\ \emph {et~al.}(2011)\citenamefont
  {Kitagawa}, \citenamefont {Oka}, \citenamefont {Brataas}, \citenamefont
  {Fu},\ and\ \citenamefont {Demler}}]{Kitagawa2011}%
  \BibitemOpen
  \bibfield  {author} {\bibinfo {author} {\bibfnamefont {Takuya}\ \bibnamefont
  {Kitagawa}}, \bibinfo {author} {\bibfnamefont {Takashi}\ \bibnamefont {Oka}},
  \bibinfo {author} {\bibfnamefont {Arne}\ \bibnamefont {Brataas}}, \bibinfo
  {author} {\bibfnamefont {Liang}\ \bibnamefont {Fu}}, \ and\ \bibinfo {author}
  {\bibfnamefont {Eugene}\ \bibnamefont {Demler}},\ }\bibfield  {title}
  {\enquote {\bibinfo {title} {Transport properties of nonequilibrium systems
  under the application of light: Photoinduced quantum hall insulators without
  landau levels},}\ }\href {\doibase 10.1103/PhysRevB.84.235108} {\bibfield
  {journal} {\bibinfo  {journal} {Phys. Rev. B}\ }\textbf {\bibinfo {volume}
  {84}},\ \bibinfo {pages} {235108} (\bibinfo {year} {2011})}\BibitemShut
  {NoStop}%
\bibitem [{\citenamefont {Delplace}\ \emph {et~al.}(2013)\citenamefont
  {Delplace}, \citenamefont {G\'omez-Le\'on},\ and\ \citenamefont
  {Platero}}]{Delplace2013}%
  \BibitemOpen
  \bibfield  {author} {\bibinfo {author} {\bibfnamefont {Pierre}\ \bibnamefont
  {Delplace}}, \bibinfo {author} {\bibfnamefont {\'Alvaro}\ \bibnamefont
  {G\'omez-Le\'on}}, \ and\ \bibinfo {author} {\bibfnamefont {Gloria}\
  \bibnamefont {Platero}},\ }\bibfield  {title} {\enquote {\bibinfo {title}
  {Merging of dirac points and floquet topological transitions in ac-driven
  graphene},}\ }\href {\doibase 10.1103/PhysRevB.88.245422} {\bibfield
  {journal} {\bibinfo  {journal} {Phys. Rev. B}\ }\textbf {\bibinfo {volume}
  {88}},\ \bibinfo {pages} {245422} (\bibinfo {year} {2013})}\BibitemShut
  {NoStop}%
\bibitem [{\citenamefont {Katan}\ and\ \citenamefont
  {Podolsky}(2013)}]{Podolsky2013}%
  \BibitemOpen
  \bibfield  {author} {\bibinfo {author} {\bibfnamefont {Yaniv~Tenenbaum}\
  \bibnamefont {Katan}}\ and\ \bibinfo {author} {\bibfnamefont {Daniel}\
  \bibnamefont {Podolsky}},\ }\bibfield  {title} {\enquote {\bibinfo {title}
  {Modulated floquet topological insulators},}\ }\href@noop {} {\bibfield
  {journal} {\bibinfo  {journal} {Phys. Rev. Lett.}\ }\textbf {\bibinfo
  {volume} {110}},\ \bibinfo {pages} {016802} (\bibinfo {year}
  {2013})}\BibitemShut {NoStop}%
\bibitem [{\citenamefont {Liu}\ \emph {et~al.}(2013)\citenamefont {Liu},
  \citenamefont {Levchenko},\ and\ \citenamefont {Baranger}}]{Liu2013}%
  \BibitemOpen
  \bibfield  {author} {\bibinfo {author} {\bibfnamefont {Dong~E.}\ \bibnamefont
  {Liu}}, \bibinfo {author} {\bibfnamefont {Alex}\ \bibnamefont {Levchenko}}, \
  and\ \bibinfo {author} {\bibfnamefont {Harold~U.}\ \bibnamefont {Baranger}},\
  }\bibfield  {title} {\enquote {\bibinfo {title} {Floquet majorana fermions
  for topological qubits in superconducting devices and cold-atom systems},}\
  }\href {\doibase 10.1103/PhysRevLett.111.047002} {\bibfield  {journal}
  {\bibinfo  {journal} {Phys. Rev. Lett.}\ }\textbf {\bibinfo {volume} {111}},\
  \bibinfo {pages} {047002} (\bibinfo {year} {2013})}\BibitemShut {NoStop}%
\bibitem [{\citenamefont {Titum}\ \emph {et~al.}(2015)\citenamefont {Titum},
  \citenamefont {Lindner}, \citenamefont {Rechtsman},\ and\ \citenamefont
  {Refael}}]{Titum2015}%
  \BibitemOpen
  \bibfield  {author} {\bibinfo {author} {\bibfnamefont {Paraj}\ \bibnamefont
  {Titum}}, \bibinfo {author} {\bibfnamefont {Netanel~H.}\ \bibnamefont
  {Lindner}}, \bibinfo {author} {\bibfnamefont {Mikael~C.}\ \bibnamefont
  {Rechtsman}}, \ and\ \bibinfo {author} {\bibfnamefont {Gil}\ \bibnamefont
  {Refael}},\ }\bibfield  {title} {\enquote {\bibinfo {title} {Disorder-induced
  floquet topological insulators},}\ }\href@noop {} {\bibfield  {journal}
  {\bibinfo  {journal} {Phys. Rev. Lett.}\ }\textbf {\bibinfo {volume} {114}},\
  \bibinfo {pages} {056801} (\bibinfo {year} {2015})}\BibitemShut {NoStop}%
\bibitem [{\citenamefont {Usaj}\ \emph {et~al.}(2014)\citenamefont {Usaj},
  \citenamefont {Perez-Piskunow}, \citenamefont {Foa~Torres},\ and\
  \citenamefont {Balseiro}}]{TorresPRB2014}%
  \BibitemOpen
  \bibfield  {author} {\bibinfo {author} {\bibfnamefont {Gonzalo}\ \bibnamefont
  {Usaj}}, \bibinfo {author} {\bibfnamefont {P.~M.}\ \bibnamefont
  {Perez-Piskunow}}, \bibinfo {author} {\bibfnamefont {L.~E.~F.}\ \bibnamefont
  {Foa~Torres}}, \ and\ \bibinfo {author} {\bibfnamefont {C.~A.}\ \bibnamefont
  {Balseiro}},\ }\bibfield  {title} {\enquote {\bibinfo {title} {Irradiated
  graphene as a tunable floquet topological insulator},}\ }\href@noop {}
  {\bibfield  {journal} {\bibinfo  {journal} {Phys. Rev. B}\ }\textbf {\bibinfo
  {volume} {90}},\ \bibinfo {pages} {115423} (\bibinfo {year}
  {2014})}\BibitemShut {NoStop}%
\bibitem [{\citenamefont {Foa~Torres}\ \emph {et~al.}(2014)\citenamefont
  {Foa~Torres}, \citenamefont {Perez-Piskunow}, \citenamefont {Balseiro},\ and\
  \citenamefont {Usaj}}]{TorresPRL2014}%
  \BibitemOpen
  \bibfield  {author} {\bibinfo {author} {\bibfnamefont {L.~E.~F.}\
  \bibnamefont {Foa~Torres}}, \bibinfo {author} {\bibfnamefont {P.~M.}\
  \bibnamefont {Perez-Piskunow}}, \bibinfo {author} {\bibfnamefont {C.~A.}\
  \bibnamefont {Balseiro}}, \ and\ \bibinfo {author} {\bibfnamefont {Gonzalo}\
  \bibnamefont {Usaj}},\ }\bibfield  {title} {\enquote {\bibinfo {title}
  {Multiterminal conductance of a floquet topological insulator},}\ }\href@noop
  {} {\bibfield  {journal} {\bibinfo  {journal} {Phys. Rev. Lett.}\ }\textbf
  {\bibinfo {volume} {113}},\ \bibinfo {pages} {266801} (\bibinfo {year}
  {2014})}\BibitemShut {NoStop}%
\bibitem [{\citenamefont {D'Alessio}\ and\ \citenamefont
  {Rigol}(2014)}]{AlessioArxiv2014}%
  \BibitemOpen
  \bibfield  {author} {\bibinfo {author} {\bibfnamefont {L.}~\bibnamefont
  {D'Alessio}}\ and\ \bibinfo {author} {\bibfnamefont {M.}~\bibnamefont
  {Rigol}},\ }\bibfield  {title} {\enquote {\bibinfo {title} {Dynamical
  preparation of floquet chern insulators:a no-go theorem, the bott index, and
  boundary effects},}\ }\href@noop {} {\bibfield  {journal} {\bibinfo
  {journal} {arXiv:1409.6319}\ } (\bibinfo {year} {2014})}\BibitemShut
  {NoStop}%
\bibitem [{\citenamefont {Dehghani}\ \emph
  {et~al.}(2014{\natexlab{a}})\citenamefont {Dehghani}, \citenamefont {Oka},\
  and\ \citenamefont {Mitra}}]{Dehghani2014}%
  \BibitemOpen
  \bibfield  {author} {\bibinfo {author} {\bibfnamefont {Hossein}\ \bibnamefont
  {Dehghani}}, \bibinfo {author} {\bibfnamefont {Takashi}\ \bibnamefont {Oka}},
  \ and\ \bibinfo {author} {\bibfnamefont {Aditi}\ \bibnamefont {Mitra}},\
  }\bibfield  {title} {\enquote {\bibinfo {title} {Dissipative floquet
  topological systems},}\ }\href@noop {} {\bibfield  {journal} {\bibinfo
  {journal} {Phys. Rev. B}\ }\textbf {\bibinfo {volume} {90}},\ \bibinfo
  {pages} {195429} (\bibinfo {year} {2014}{\natexlab{a}})}\BibitemShut
  {NoStop}%
\bibitem [{\citenamefont {Dehghani}\ \emph
  {et~al.}(2014{\natexlab{b}})\citenamefont {Dehghani}, \citenamefont {Oka},\
  and\ \citenamefont {Mitra}}]{Dehghani2014b}%
  \BibitemOpen
  \bibfield  {author} {\bibinfo {author} {\bibfnamefont {Hossein}\ \bibnamefont
  {Dehghani}}, \bibinfo {author} {\bibfnamefont {Takashi}\ \bibnamefont {Oka}},
  \ and\ \bibinfo {author} {\bibfnamefont {Aditi}\ \bibnamefont {Mitra}},\
  }\bibfield  {title} {\enquote {\bibinfo {title} {Out of equilibrium electrons
  and the hall conductance of a floquet topological insulator},}\ }\href@noop
  {} {\bibfield  {journal} {\bibinfo  {journal} {arXiv:1412.8469}\ } (\bibinfo
  {year} {2014}{\natexlab{b}})}\BibitemShut {NoStop}%
\bibitem [{\citenamefont {Sentef}\ \emph {et~al.}(2015)\citenamefont {Sentef},
  \citenamefont {Claassen}, \citenamefont {Kemper}, \citenamefont {Moritz},
  \citenamefont {Oka},\ and\ \citenamefont {Freericks}}]{Sentef2015}%
  \BibitemOpen
  \bibfield  {author} {\bibinfo {author} {\bibfnamefont {M.A.}\ \bibnamefont
  {Sentef}}, \bibinfo {author} {\bibfnamefont {M.}~\bibnamefont {Claassen}},
  \bibinfo {author} {\bibfnamefont {A.F.}\ \bibnamefont {Kemper}}, \bibinfo
  {author} {\bibfnamefont {B.}~\bibnamefont {Moritz}}, \bibinfo {author}
  {\bibfnamefont {T.}~\bibnamefont {Oka}}, \ and\ \bibinfo {author}
  {\bibfnamefont {T.P}\ \bibnamefont {Freericks}, \bibfnamefont
  {J.Kand~Devereaux}},\ }\bibfield  {title} {\enquote {\bibinfo {title} {Theory
  of floquet band formation and local pseudospin textures in pump-probe
  photoemission of graphene},}\ }\href@noop {} {\bibfield  {journal} {\bibinfo
  {journal} {Nat. Comm.}\ }\textbf {\bibinfo {volume} {6}},\ \bibinfo {pages}
  {7047} (\bibinfo {year} {2015})}\BibitemShut {NoStop}%
\bibitem [{\citenamefont {Seetharam}\ \emph {et~al.}(2015)\citenamefont
  {Seetharam}, \citenamefont {Bardyn}, \citenamefont {Lindner}, \citenamefont
  {Rudner},\ and\ \citenamefont {Refael}}]{Seetharam2015}%
  \BibitemOpen
  \bibfield  {author} {\bibinfo {author} {\bibfnamefont {Karthik~I.}\
  \bibnamefont {Seetharam}}, \bibinfo {author} {\bibfnamefont
  {Charles-Edouard}\ \bibnamefont {Bardyn}}, \bibinfo {author} {\bibfnamefont
  {Netanel~H.}\ \bibnamefont {Lindner}}, \bibinfo {author} {\bibfnamefont
  {Mark~S.}\ \bibnamefont {Rudner}}, \ and\ \bibinfo {author} {\bibfnamefont
  {Gil}\ \bibnamefont {Refael}},\ }\bibfield  {title} {\enquote {\bibinfo
  {title} {Controlled population of floquet-bloch states via coupling to bose
  and fermi baths},}\ }\href@noop {} {\bibfield  {journal} {\bibinfo  {journal}
  {Phys. Rev. X}\ }\textbf {\bibinfo {volume} {5}},\ \bibinfo {pages} {041050}
  (\bibinfo {year} {2015})}\BibitemShut {NoStop}%
\bibitem [{\citenamefont {Iadecola}\ \emph {et~al.}(2015)\citenamefont
  {Iadecola}, \citenamefont {Neupert},\ and\ \citenamefont
  {Chamon}}]{Iadecola2015}%
  \BibitemOpen
  \bibfield  {author} {\bibinfo {author} {\bibfnamefont {Thomas}\ \bibnamefont
  {Iadecola}}, \bibinfo {author} {\bibfnamefont {Titus}\ \bibnamefont
  {Neupert}}, \ and\ \bibinfo {author} {\bibfnamefont {Claudio}\ \bibnamefont
  {Chamon}},\ }\bibfield  {title} {\enquote {\bibinfo {title} {Occupation of
  topological floquet bands in open systems},}\ }\href@noop {} {\bibfield
  {journal} {\bibinfo  {journal} {Phys. Rev. B}\ }\textbf {\bibinfo {volume}
  {91}},\ \bibinfo {pages} {235133} (\bibinfo {year} {2015})}\BibitemShut
  {NoStop}%
\bibitem [{\citenamefont {{Klinovaja}}\ \emph {et~al.}(2015)\citenamefont
  {{Klinovaja}}, \citenamefont {{Stano}},\ and\ \citenamefont
  {{Loss}}}]{Klinovaja2015}%
  \BibitemOpen
  \bibfield  {author} {\bibinfo {author} {\bibfnamefont {J.}~\bibnamefont
  {{Klinovaja}}}, \bibinfo {author} {\bibfnamefont {P.}~\bibnamefont
  {{Stano}}}, \ and\ \bibinfo {author} {\bibfnamefont {D.}~\bibnamefont
  {{Loss}}},\ }\bibfield  {title} {\enquote {\bibinfo {title} {{Topological
  Floquet Phases in Driven Coupled Rashba Nanowires}},}\ }\href@noop {}
  {\bibfield  {journal} {\bibinfo  {journal} {arXiv:1510.03640}\ } (\bibinfo
  {year} {2015})}\BibitemShut {NoStop}%
\bibitem [{\citenamefont {{Gannot}}(2015)}]{Gannot2015}%
  \BibitemOpen
  \bibfield  {author} {\bibinfo {author} {\bibfnamefont {Y.}~\bibnamefont
  {{Gannot}}},\ }\bibfield  {title} {\enquote {\bibinfo {title} {{Effects of
  Disorder on a 1-D Floquet Symmetry Protected Topological Phase}},}\
  }\href@noop {} {\bibfield  {journal} {\bibinfo  {journal} {arXiv:1512.04190}\
  } (\bibinfo {year} {2015})}\BibitemShut {NoStop}%
\bibitem [{\citenamefont {Wang}\ \emph {et~al.}(2013)\citenamefont {Wang},
  \citenamefont {Steinberg}, \citenamefont {Jarillo-Herrero},\ and\
  \citenamefont {Gedik}}]{Wang2013}%
  \BibitemOpen
  \bibfield  {author} {\bibinfo {author} {\bibfnamefont {Y.~H.}\ \bibnamefont
  {Wang}}, \bibinfo {author} {\bibfnamefont {H.}~\bibnamefont {Steinberg}},
  \bibinfo {author} {\bibfnamefont {P.}~\bibnamefont {Jarillo-Herrero}}, \ and\
  \bibinfo {author} {\bibfnamefont {N.}~\bibnamefont {Gedik}},\ }\bibfield
  {title} {\enquote {\bibinfo {title} {Observation of floquet-bloch states on
  the surface of a topological insulator},}\ }\href {\doibase
  10.1126/science.1239834} {\bibfield  {journal} {\bibinfo  {journal}
  {Science}\ }\textbf {\bibinfo {volume} {342}},\ \bibinfo {pages} {453--457}
  (\bibinfo {year} {2013})}\BibitemShut {NoStop}%
\bibitem [{\citenamefont {Eckardt}\ and\ \citenamefont
  {Anisimovas}(2015{\natexlab{a}})}]{Gedik2016}%
  \BibitemOpen
  \bibfield  {author} {\bibinfo {author} {\bibfnamefont {Andre}\ \bibnamefont
  {Eckardt}}\ and\ \bibinfo {author} {\bibfnamefont {Egidijus}\ \bibnamefont
  {Anisimovas}},\ }\bibfield  {title} {\enquote {\bibinfo {title}
  {High-frequency approximation for periodically driven quantum systems from a
  floquet-space perspective},}\ }\href@noop {} {\bibfield  {journal} {\bibinfo
  {journal} {New Journal of Physics}\ }\textbf {\bibinfo {volume} {17}},\
  \bibinfo {pages} {093039} (\bibinfo {year} {2015}{\natexlab{a}})}\BibitemShut
  {NoStop}%
\bibitem [{\citenamefont {Jotzu}\ \emph {et~al.}(2014)\citenamefont {Jotzu},
  \citenamefont {Messer}, \citenamefont {Desbuquois}, \citenamefont {Lebrat},
  \citenamefont {Uehlinger}, \citenamefont {Greif},\ and\ \citenamefont
  {Esslinger}}]{Jotzu2014}%
  \BibitemOpen
  \bibfield  {author} {\bibinfo {author} {\bibfnamefont {Gregor}\ \bibnamefont
  {Jotzu}}, \bibinfo {author} {\bibfnamefont {Michael}\ \bibnamefont {Messer}},
  \bibinfo {author} {\bibfnamefont {R\'emi}\ \bibnamefont {Desbuquois}},
  \bibinfo {author} {\bibfnamefont {Martin}\ \bibnamefont {Lebrat}}, \bibinfo
  {author} {\bibfnamefont {Thomas}\ \bibnamefont {Uehlinger}}, \bibinfo
  {author} {\bibfnamefont {Daniel}\ \bibnamefont {Greif}}, \ and\ \bibinfo
  {author} {\bibfnamefont {Tilman}\ \bibnamefont {Esslinger}},\ }\bibfield
  {title} {\enquote {\bibinfo {title} {Experimental realization of the
  topological haldane model with ultracold fermions},}\ }\href {\doibase
  10.1038/nature13915} {\bibfield  {journal} {\bibinfo  {journal} {Nature}\
  }\textbf {\bibinfo {volume} {515}},\ \bibinfo {pages} {237--240} (\bibinfo
  {year} {2014})}\BibitemShut {NoStop}%
\bibitem [{\citenamefont {Rechtsman}\ \emph {et~al.}(2013)\citenamefont
  {Rechtsman}, \citenamefont {Zeuner}, \citenamefont {Plotnik}, \citenamefont
  {Lumer}, \citenamefont {Podolsky}, \citenamefont {Dreisow}, \citenamefont
  {Nolte}, \citenamefont {Segev},\ and\ \citenamefont
  {Szameit}}]{Rechtsman2013}%
  \BibitemOpen
  \bibfield  {author} {\bibinfo {author} {\bibfnamefont {M.~C.}\ \bibnamefont
  {Rechtsman}}, \bibinfo {author} {\bibfnamefont {J.~M.}\ \bibnamefont
  {Zeuner}}, \bibinfo {author} {\bibfnamefont {Y.}~\bibnamefont {Plotnik}},
  \bibinfo {author} {\bibfnamefont {Y.}~\bibnamefont {Lumer}}, \bibinfo
  {author} {\bibfnamefont {D.}~\bibnamefont {Podolsky}}, \bibinfo {author}
  {\bibfnamefont {F.}~\bibnamefont {Dreisow}}, \bibinfo {author} {\bibfnamefont
  {S.}~\bibnamefont {Nolte}}, \bibinfo {author} {\bibfnamefont
  {M.}~\bibnamefont {Segev}}, \ and\ \bibinfo {author} {\bibfnamefont
  {A.}~\bibnamefont {Szameit}},\ }\bibfield  {title} {\enquote {\bibinfo
  {title} {Photonic floquet topological insulators},}\ }\href {\doibase
  10.1038/nature12066} {\bibfield  {journal} {\bibinfo  {journal} {Nature}\
  }\textbf {\bibinfo {volume} {496}},\ \bibinfo {pages} {196--200} (\bibinfo
  {year} {2013})}\BibitemShut {NoStop}%
\bibitem [{\citenamefont {Hu}\ \emph {et~al.}(2015)\citenamefont {Hu},
  \citenamefont {Pillay}, \citenamefont {Wu}, \citenamefont {Pasek},
  \citenamefont {Shum},\ and\ \citenamefont {Chong}}]{Hu2015}%
  \BibitemOpen
  \bibfield  {author} {\bibinfo {author} {\bibfnamefont {Wenchao}\ \bibnamefont
  {Hu}}, \bibinfo {author} {\bibfnamefont {Jason~C.}\ \bibnamefont {Pillay}},
  \bibinfo {author} {\bibfnamefont {Kan}\ \bibnamefont {Wu}}, \bibinfo {author}
  {\bibfnamefont {Michael}\ \bibnamefont {Pasek}}, \bibinfo {author}
  {\bibfnamefont {Perry~Ping}\ \bibnamefont {Shum}}, \ and\ \bibinfo {author}
  {\bibfnamefont {Y.~D.}\ \bibnamefont {Chong}},\ }\bibfield  {title} {\enquote
  {\bibinfo {title} {Measurement of a topological edge invariant in a microwave
  network},}\ }\href@noop {} {\bibfield  {journal} {\bibinfo  {journal} {Phys.
  Rev. X}\ }\textbf {\bibinfo {volume} {5}},\ \bibinfo {pages} {011012}
  (\bibinfo {year} {2015})}\BibitemShut {NoStop}%
\bibitem [{\citenamefont {Lazarides}\ \emph
  {et~al.}(2014{\natexlab{a}})\citenamefont {Lazarides}, \citenamefont {Das},\
  and\ \citenamefont {Moessner}}]{LazaridesDasMoessner2014}%
  \BibitemOpen
  \bibfield  {author} {\bibinfo {author} {\bibfnamefont {Achilleas}\
  \bibnamefont {Lazarides}}, \bibinfo {author} {\bibfnamefont {Arnab}\
  \bibnamefont {Das}}, \ and\ \bibinfo {author} {\bibfnamefont {Roderich}\
  \bibnamefont {Moessner}},\ }\bibfield  {title} {\enquote {\bibinfo {title}
  {Equilibrium states of generic quantum systems subject to periodic
  driving},}\ }\href@noop {} {\bibfield  {journal} {\bibinfo  {journal} {Phys.
  Rev. E}\ }\textbf {\bibinfo {volume} {90}},\ \bibinfo {pages} {012110}
  (\bibinfo {year} {2014}{\natexlab{a}})}\BibitemShut {NoStop}%
\bibitem [{\citenamefont {DAlessio}\ and\ \citenamefont
  {Rigol}(2014)}]{DAlessio2014}%
  \BibitemOpen
  \bibfield  {author} {\bibinfo {author} {\bibfnamefont {Luca}\ \bibnamefont
  {DAlessio}}\ and\ \bibinfo {author} {\bibfnamefont {Marcos}\ \bibnamefont
  {Rigol}},\ }\bibfield  {title} {\enquote {\bibinfo {title} {Long-time
  behavior of isolated periodically driven interacting lattice systems},}\
  }\href@noop {} {\bibfield  {journal} {\bibinfo  {journal} {Phys. Rev. X}\
  }\textbf {\bibinfo {volume} {4}},\ \bibinfo {pages} {041048} (\bibinfo {year}
  {2014})}\BibitemShut {NoStop}%
\bibitem [{\citenamefont {Lazarides}\ \emph {et~al.}(2015)\citenamefont
  {Lazarides}, \citenamefont {Das},\ and\ \citenamefont
  {Moessner}}]{Lazarides2015}%
  \BibitemOpen
  \bibfield  {author} {\bibinfo {author} {\bibfnamefont {Achilleas}\
  \bibnamefont {Lazarides}}, \bibinfo {author} {\bibfnamefont {Arnab}\
  \bibnamefont {Das}}, \ and\ \bibinfo {author} {\bibfnamefont {Roderich}\
  \bibnamefont {Moessner}},\ }\bibfield  {title} {\enquote {\bibinfo {title}
  {Fate of many-body localization under periodic driving},}\ }\href {\doibase
  10.1103/PhysRevLett.115.030402} {\bibfield  {journal} {\bibinfo  {journal}
  {Phys. Rev. Lett.}\ }\textbf {\bibinfo {volume} {115}},\ \bibinfo {pages}
  {030402} (\bibinfo {year} {2015})}\BibitemShut {NoStop}%
\bibitem [{\citenamefont {Citro}\ \emph {et~al.}(2015)\citenamefont {Citro},
  \citenamefont {Torre}, \citenamefont {D’Alessio}, \citenamefont
  {Polkovnikov}, \citenamefont {Babadi}, \citenamefont {Oka},\ and\
  \citenamefont {Demler}}]{Citro2015}%
  \BibitemOpen
  \bibfield  {author} {\bibinfo {author} {\bibfnamefont {Roberta}\ \bibnamefont
  {Citro}}, \bibinfo {author} {\bibfnamefont {Emanuele G.~Dalla}\ \bibnamefont
  {Torre}}, \bibinfo {author} {\bibfnamefont {Luca}\ \bibnamefont
  {D’Alessio}}, \bibinfo {author} {\bibfnamefont {Anatoli}\ \bibnamefont
  {Polkovnikov}}, \bibinfo {author} {\bibfnamefont {Mehrtash}\ \bibnamefont
  {Babadi}}, \bibinfo {author} {\bibfnamefont {Takashi}\ \bibnamefont {Oka}}, \
  and\ \bibinfo {author} {\bibfnamefont {Eugene}\ \bibnamefont {Demler}},\
  }\bibfield  {title} {\enquote {\bibinfo {title} {Dynamical stability of a
  many-body kapitza pendulum},}\ }\href@noop {} {\bibfield  {journal} {\bibinfo
   {journal} {Annals of Physics}\ }\textbf {\bibinfo {volume} {360}},\ \bibinfo
  {pages} {694 -- 710} (\bibinfo {year} {2015})}\BibitemShut {NoStop}%
\bibitem [{\citenamefont {Chandran}\ and\ \citenamefont
  {Sondhi}(2015)}]{Chandran2015}%
  \BibitemOpen
  \bibfield  {author} {\bibinfo {author} {\bibfnamefont {Anushya}\ \bibnamefont
  {Chandran}}\ and\ \bibinfo {author} {\bibfnamefont {S.~L.}\ \bibnamefont
  {Sondhi}},\ }\bibfield  {title} {\enquote {\bibinfo {title} {Interaction
  stabilized steady states in the driven o(n) model},}\ }\href@noop {}
  {\bibfield  {journal} {\bibinfo  {journal} {arXiv:1506.08836}\ } (\bibinfo
  {year} {2015})}\BibitemShut {NoStop}%
\bibitem [{\citenamefont {{Kukuljan}}\ and\ \citenamefont
  {{Prosen}}(2015)}]{Prozen}%
  \BibitemOpen
  \bibfield  {author} {\bibinfo {author} {\bibfnamefont {I.}~\bibnamefont
  {{Kukuljan}}}\ and\ \bibinfo {author} {\bibfnamefont {T.}~\bibnamefont
  {{Prosen}}},\ }\bibfield  {title} {\enquote {\bibinfo {title} {Corner
  transfer matrices for 2d strongly coupled many-body floquet systems},}\
  }\href@noop {} {\bibfield  {journal} {\bibinfo  {journal} {arXiv:1512.06601}\
  } (\bibinfo {year} {2015})}\BibitemShut {NoStop}%
\bibitem [{\citenamefont {Lazarides}\ \emph
  {et~al.}(2014{\natexlab{b}})\citenamefont {Lazarides}, \citenamefont {Das},\
  and\ \citenamefont {Moessner}}]{Lazarides2014}%
  \BibitemOpen
  \bibfield  {author} {\bibinfo {author} {\bibfnamefont {Achilleas}\
  \bibnamefont {Lazarides}}, \bibinfo {author} {\bibfnamefont {Arnab}\
  \bibnamefont {Das}}, \ and\ \bibinfo {author} {\bibfnamefont {Roderich}\
  \bibnamefont {Moessner}},\ }\bibfield  {title} {\enquote {\bibinfo {title}
  {Periodic thermodynamics of isolated quantum systems},}\ }\href {\doibase
  10.1103/PhysRevLett.112.150401} {\bibfield  {journal} {\bibinfo  {journal}
  {Phys. Rev. Lett.}\ }\textbf {\bibinfo {volume} {112}},\ \bibinfo {pages}
  {150401} (\bibinfo {year} {2014}{\natexlab{b}})}\BibitemShut {NoStop}%
\bibitem [{\citenamefont {Ponte}\ \emph {et~al.}(2014)\citenamefont {Ponte},
  \citenamefont {Chandran}, \citenamefont {Papi\'c},\ and\ \citenamefont
  {Abanin}}]{Ponte2014}%
  \BibitemOpen
  \bibfield  {author} {\bibinfo {author} {\bibfnamefont {Pedro}\ \bibnamefont
  {Ponte}}, \bibinfo {author} {\bibfnamefont {Anushya}\ \bibnamefont
  {Chandran}}, \bibinfo {author} {\bibfnamefont {Z.}~\bibnamefont {Papi\'c}}, \
  and\ \bibinfo {author} {\bibfnamefont {Dmitry~A.}\ \bibnamefont {Abanin}},\
  }\bibfield  {title} {\enquote {\bibinfo {title} {Periodically driven ergodic
  and many-body localized quantum systems},}\ }\href {\doibase
  10.1016/j.aop.2014.11.008} {\bibfield  {journal} {\bibinfo  {journal} {Annals
  of Physics}\ }\textbf {\bibinfo {volume} {353}},\ \bibinfo {pages} {196--204}
  (\bibinfo {year} {2014})}\BibitemShut {NoStop}%
\bibitem [{\citenamefont {Abanin}\ \emph {et~al.}(2014)\citenamefont {Abanin},
  \citenamefont {De~Roeck},\ and\ \citenamefont {Huveneers}}]{Abanin2014}%
  \BibitemOpen
  \bibfield  {author} {\bibinfo {author} {\bibfnamefont {D.}~\bibnamefont
  {Abanin}}, \bibinfo {author} {\bibfnamefont {W.}~\bibnamefont {De~Roeck}}, \
  and\ \bibinfo {author} {\bibfnamefont {F.}~\bibnamefont {Huveneers}},\
  }\bibfield  {title} {\enquote {\bibinfo {title} {A theory of many-body
  localization in periodically driven systems},}\ }\href@noop {} {\bibfield
  {journal} {\bibinfo  {journal} {arXiv:1412.2752}\ } (\bibinfo {year}
  {2014})}\BibitemShut {NoStop}%
\bibitem [{\citenamefont {{Khemani}}\ \emph {et~al.}(2015)\citenamefont
  {{Khemani}}, \citenamefont {{Lazarides}}, \citenamefont {{Moessner}},\ and\
  \citenamefont {{Sondhi}}}]{Khemani2015}%
  \BibitemOpen
  \bibfield  {author} {\bibinfo {author} {\bibfnamefont {V.}~\bibnamefont
  {{Khemani}}}, \bibinfo {author} {\bibfnamefont {A.}~\bibnamefont
  {{Lazarides}}}, \bibinfo {author} {\bibfnamefont {R.}~\bibnamefont
  {{Moessner}}}, \ and\ \bibinfo {author} {\bibfnamefont {S.~L.}\ \bibnamefont
  {{Sondhi}}},\ }\bibfield  {title} {\enquote {\bibinfo {title} {{On the phase
  structure of driven quantum systems}},}\ }\href@noop {} {\bibfield  {journal}
  {\bibinfo  {journal} {arXiv:1508.03344}\ } (\bibinfo {year}
  {2015})}\BibitemShut {NoStop}%
\bibitem [{\citenamefont {von Keyserlingk}\ and\ \citenamefont
  {Sondhi}(2016{\natexlab{a}})}]{vonKeyserlingk2016a}%
  \BibitemOpen
  \bibfield  {author} {\bibinfo {author} {\bibfnamefont {C.~W.}\ \bibnamefont
  {von Keyserlingk}}\ and\ \bibinfo {author} {\bibfnamefont {S.~L.}\
  \bibnamefont {Sondhi}},\ }\bibfield  {title} {\enquote {\bibinfo {title}
  {Phase structure of 1d interacting floquet systems i: Abelian spts},}\
  }\href@noop {} {\bibfield  {journal} {\bibinfo  {journal} {arXiv:1602.02157}\
  } (\bibinfo {year} {2016}{\natexlab{a}})}\BibitemShut {NoStop}%
\bibitem [{\citenamefont {von Keyserlingk}\ and\ \citenamefont
  {Sondhi}(2016{\natexlab{b}})}]{vonKeyserlingk2016b}%
  \BibitemOpen
  \bibfield  {author} {\bibinfo {author} {\bibfnamefont {C.~W.}\ \bibnamefont
  {von Keyserlingk}}\ and\ \bibinfo {author} {\bibfnamefont {S.~L.}\
  \bibnamefont {Sondhi}},\ }\bibfield  {title} {\enquote {\bibinfo {title} {1d
  many-body localized floquet systems ii: Symmetry-broken phases},}\
  }\href@noop {} {\bibfield  {journal} {\bibinfo  {journal} {arXiv:1602.06949}\
  } (\bibinfo {year} {2016}{\natexlab{b}})}\BibitemShut {NoStop}%
\bibitem [{\citenamefont {Potter}\ \emph {et~al.}(2016)\citenamefont {Potter},
  \citenamefont {Morimoto},\ and\ \citenamefont {Vishwanath}}]{Potter2016}%
  \BibitemOpen
  \bibfield  {author} {\bibinfo {author} {\bibfnamefont {Andrew~C.}\
  \bibnamefont {Potter}}, \bibinfo {author} {\bibfnamefont {Takahiro}\
  \bibnamefont {Morimoto}}, \ and\ \bibinfo {author} {\bibfnamefont {Ashvin}\
  \bibnamefont {Vishwanath}},\ }\bibfield  {title} {\enquote {\bibinfo {title}
  {Topological classification of interacting 1d floquet phases},}\ }\href@noop
  {} {\bibfield  {journal} {\bibinfo  {journal} {arXiv:1602.05194}\ } (\bibinfo
  {year} {2016})}\BibitemShut {NoStop}%
\bibitem [{\citenamefont {Else}\ and\ \citenamefont {Nayak}(2016)}]{Nayak2016}%
  \BibitemOpen
  \bibfield  {author} {\bibinfo {author} {\bibfnamefont {Dominic~V.}\
  \bibnamefont {Else}}\ and\ \bibinfo {author} {\bibfnamefont {Chetan}\
  \bibnamefont {Nayak}},\ }\bibfield  {title} {\enquote {\bibinfo {title} {On
  the classification of topological phases in periodically driven interacting
  systems},}\ }\href@noop {} {\bibfield  {journal} {\bibinfo  {journal}
  {arXiv:1602.04804}\ } (\bibinfo {year} {2016})}\BibitemShut {NoStop}%
\bibitem [{\citenamefont {Roy}\ and\ \citenamefont {Harper}(2016)}]{Roy2016}%
  \BibitemOpen
  \bibfield  {author} {\bibinfo {author} {\bibfnamefont {Rahul}\ \bibnamefont
  {Roy}}\ and\ \bibinfo {author} {\bibfnamefont {Fenner}\ \bibnamefont
  {Harper}},\ }\bibfield  {title} {\enquote {\bibinfo {title} {Abelian floquet
  spt phases in 1d},}\ }\href@noop {} {\bibfield  {journal} {\bibinfo
  {journal} {arXiv:1602.08089}\ } (\bibinfo {year} {2016})}\BibitemShut
  {NoStop}%
\bibitem [{\citenamefont {Abanin}\ \emph {et~al.}(2015)\citenamefont {Abanin},
  \citenamefont {De~Roeck},\ and\ \citenamefont {Huveneers}}]{AbaninHighFreq1}%
  \BibitemOpen
  \bibfield  {author} {\bibinfo {author} {\bibfnamefont {Dmitry~A.}\
  \bibnamefont {Abanin}}, \bibinfo {author} {\bibfnamefont {Wojciech}\
  \bibnamefont {De~Roeck}}, \ and\ \bibinfo {author} {\bibfnamefont
  {Fran\ifmmode \mbox{\c{c}}\else~\c{c}\fi{}ois}\ \bibnamefont {Huveneers}},\
  }\bibfield  {title} {\enquote {\bibinfo {title} {Exponentially slow heating
  in periodically driven many-body systems},}\ }\href@noop {} {\bibfield
  {journal} {\bibinfo  {journal} {Phys. Rev. Lett.}\ }\textbf {\bibinfo
  {volume} {115}},\ \bibinfo {pages} {256803} (\bibinfo {year}
  {2015})}\BibitemShut {NoStop}%
\bibitem [{\citenamefont {{Abanin}}\ \emph {et~al.}(2015)\citenamefont
  {{Abanin}}, \citenamefont {{De Roeck}},\ and\ \citenamefont
  {{Ho}}}]{AbaninHighFreq2}%
  \BibitemOpen
  \bibfield  {author} {\bibinfo {author} {\bibfnamefont {D.~A.}\ \bibnamefont
  {{Abanin}}}, \bibinfo {author} {\bibfnamefont {W.}~\bibnamefont {{De
  Roeck}}}, \ and\ \bibinfo {author} {\bibfnamefont {W.~W.}\ \bibnamefont
  {{Ho}}},\ }\bibfield  {title} {\enquote {\bibinfo {title} {{Effective
  Hamiltonians, prethermalization and slow energy absorption in periodically
  driven many-body systems}},}\ }\href@noop {} {\bibfield  {journal} {\bibinfo
  {journal} {arXiv:1510.03405}\ } (\bibinfo {year} {2015})}\BibitemShut
  {NoStop}%
\bibitem [{\citenamefont {Eckardt}\ and\ \citenamefont
  {Anisimovas}(2015{\natexlab{b}})}]{Eckardt2015}%
  \BibitemOpen
  \bibfield  {author} {\bibinfo {author} {\bibfnamefont {Andre}\ \bibnamefont
  {Eckardt}}\ and\ \bibinfo {author} {\bibfnamefont {Egidijus}\ \bibnamefont
  {Anisimovas}},\ }\bibfield  {title} {\enquote {\bibinfo {title}
  {High-frequency approximation for periodically driven quantum systems from a
  floquet-space perspective},}\ }\href@noop {} {\bibfield  {journal} {\bibinfo
  {journal} {New Journal of Physics}\ }\textbf {\bibinfo {volume} {17}},\
  \bibinfo {pages} {093039} (\bibinfo {year} {2015}{\natexlab{b}})}\BibitemShut
  {NoStop}%
\bibitem [{\citenamefont {Bukov}\ \emph
  {et~al.}(2015{\natexlab{a}})\citenamefont {Bukov}, \citenamefont {Heyl},
  \citenamefont {Huse},\ and\ \citenamefont {Polkovnikov}}]{Bukov2015}%
  \BibitemOpen
  \bibfield  {author} {\bibinfo {author} {\bibfnamefont {M.}~\bibnamefont
  {Bukov}}, \bibinfo {author} {\bibfnamefont {M.}~\bibnamefont {Heyl}},
  \bibinfo {author} {\bibfnamefont {D.~A.}\ \bibnamefont {Huse}}, \ and\
  \bibinfo {author} {\bibfnamefont {A.}~\bibnamefont {Polkovnikov}},\
  }\bibfield  {title} {\enquote {\bibinfo {title} {{Heating and Many-Body
  Resonances in a Periodically-Driven Two-Band System}},}\ }\href@noop {}
  {\bibfield  {journal} {\bibinfo  {journal} {arXiv:1512.02119}\ } (\bibinfo
  {year} {2015}{\natexlab{a}})}\BibitemShut {NoStop}%
\bibitem [{\citenamefont {Berges}\ \emph {et~al.}(2004)\citenamefont {Berges},
  \citenamefont {Bors\'anyi},\ and\ \citenamefont {Wetterich}}]{Berges2004}%
  \BibitemOpen
  \bibfield  {author} {\bibinfo {author} {\bibfnamefont {J.}~\bibnamefont
  {Berges}}, \bibinfo {author} {\bibfnamefont {Sz.}\ \bibnamefont
  {Bors\'anyi}}, \ and\ \bibinfo {author} {\bibfnamefont {C.}~\bibnamefont
  {Wetterich}},\ }\bibfield  {title} {\enquote {\bibinfo {title}
  {Prethermalization},}\ }\href@noop {} {\bibfield  {journal} {\bibinfo
  {journal} {Phys. Rev. Lett.}\ }\textbf {\bibinfo {volume} {93}},\ \bibinfo
  {pages} {142002} (\bibinfo {year} {2004})}\BibitemShut {NoStop}%
\bibitem [{\citenamefont {Eckstein}\ \emph {et~al.}(2009)\citenamefont
  {Eckstein}, \citenamefont {Kollar},\ and\ \citenamefont
  {Werner}}]{Eckstein2009}%
  \BibitemOpen
  \bibfield  {author} {\bibinfo {author} {\bibfnamefont {Martin}\ \bibnamefont
  {Eckstein}}, \bibinfo {author} {\bibfnamefont {Marcus}\ \bibnamefont
  {Kollar}}, \ and\ \bibinfo {author} {\bibfnamefont {Philipp}\ \bibnamefont
  {Werner}},\ }\bibfield  {title} {\enquote {\bibinfo {title} {Thermalization
  after an interaction quench in the hubbard model},}\ }\href@noop {}
  {\bibfield  {journal} {\bibinfo  {journal} {Phys. Rev. Lett.}\ ,\ \bibinfo
  {pages} {056403}} (\bibinfo {year} {2009})}\BibitemShut {NoStop}%
\bibitem [{\citenamefont {Moeckel}\ and\ \citenamefont
  {Kehrein}(2010)}]{Moeckel2010}%
  \BibitemOpen
  \bibfield  {author} {\bibinfo {author} {\bibfnamefont {Michael}\ \bibnamefont
  {Moeckel}}\ and\ \bibinfo {author} {\bibfnamefont {Stefan}\ \bibnamefont
  {Kehrein}},\ }\bibfield  {title} {\enquote {\bibinfo {title} {Crossover from
  adiabatic to sudden interaction quenches in the hubbard model:
  prethermalization and non-equilibrium dynamics},}\ }\href@noop {} {\bibfield
  {journal} {\bibinfo  {journal} {New Journal of Physics}\ }\textbf {\bibinfo
  {volume} {12}},\ \bibinfo {pages} {055016} (\bibinfo {year}
  {2010})}\BibitemShut {NoStop}%
\bibitem [{\citenamefont {Mathey}\ and\ \citenamefont
  {Polkovnikov}(2010)}]{Mathey2010}%
  \BibitemOpen
  \bibfield  {author} {\bibinfo {author} {\bibfnamefont {L.}~\bibnamefont
  {Mathey}}\ and\ \bibinfo {author} {\bibfnamefont {A.}~\bibnamefont
  {Polkovnikov}},\ }\bibfield  {title} {\enquote {\bibinfo {title} {Light cone
  dynamics and reverse kibble-zurek mechanism in two-dimensional superfluids
  following a quantum quench},}\ }\href@noop {} {\bibfield  {journal} {\bibinfo
   {journal} {Phys. Rev. A}\ }\textbf {\bibinfo {volume} {81}},\ \bibinfo
  {pages} {033605} (\bibinfo {year} {2010})}\BibitemShut {NoStop}%
\bibitem [{\citenamefont {Bukov}\ \emph
  {et~al.}(2015{\natexlab{b}})\citenamefont {Bukov}, \citenamefont
  {Gopalakrishnan}, \citenamefont {Knap},\ and\ \citenamefont
  {Demler}}]{BukovPrethermal}%
  \BibitemOpen
  \bibfield  {author} {\bibinfo {author} {\bibfnamefont {Marin}\ \bibnamefont
  {Bukov}}, \bibinfo {author} {\bibfnamefont {Sarang}\ \bibnamefont
  {Gopalakrishnan}}, \bibinfo {author} {\bibfnamefont {Michael}\ \bibnamefont
  {Knap}}, \ and\ \bibinfo {author} {\bibfnamefont {Eugene}\ \bibnamefont
  {Demler}},\ }\bibfield  {title} {\enquote {\bibinfo {title} {Prethermal
  floquet steady states and instabilities in the periodically driven, weakly
  interacting bose-hubbard model},}\ }\href@noop {} {\bibfield  {journal}
  {\bibinfo  {journal} {Phys. Rev. Lett.}\ }\textbf {\bibinfo {volume} {115}},\
  \bibinfo {pages} {205301} (\bibinfo {year} {2015}{\natexlab{b}})}\BibitemShut
  {NoStop}%
\bibitem [{\citenamefont {Canovi}\ \emph {et~al.}(2016)\citenamefont {Canovi},
  \citenamefont {Kollar},\ and\ \citenamefont {Eckstein}}]{Eckstein2015}%
  \BibitemOpen
  \bibfield  {author} {\bibinfo {author} {\bibfnamefont {Elena}\ \bibnamefont
  {Canovi}}, \bibinfo {author} {\bibfnamefont {Marcus}\ \bibnamefont {Kollar}},
  \ and\ \bibinfo {author} {\bibfnamefont {Martin}\ \bibnamefont {Eckstein}},\
  }\bibfield  {title} {\enquote {\bibinfo {title} {Stroboscopic
  prethermalization in weakly interacting periodically driven systems},}\
  }\href@noop {} {\bibfield  {journal} {\bibinfo  {journal}
  {arXiv:1507.00991v3}\ } (\bibinfo {year} {2016})}\BibitemShut {NoStop}%
\bibitem [{\citenamefont {Kuwahara}\ \emph {et~al.}(2016)\citenamefont
  {Kuwahara}, \citenamefont {Mori},\ and\ \citenamefont
  {Saito}}]{Kuwahara2016}%
  \BibitemOpen
  \bibfield  {author} {\bibinfo {author} {\bibfnamefont {Tomotaka}\
  \bibnamefont {Kuwahara}}, \bibinfo {author} {\bibfnamefont {Takashi}\
  \bibnamefont {Mori}}, \ and\ \bibinfo {author} {\bibfnamefont {Keiji}\
  \bibnamefont {Saito}},\ }\bibfield  {title} {\enquote {\bibinfo {title}
  {Floquet–magnus theory and generic transient dynamics in periodically
  driven many-body quantum systems},}\ }\href@noop {} {\bibfield  {journal}
  {\bibinfo  {journal} {Annals of Physics}\ }\textbf {\bibinfo {volume}
  {367}},\ \bibinfo {pages} {96 -- 124} (\bibinfo {year} {2016})}\BibitemShut
  {NoStop}%
\bibitem [{\citenamefont {Mori}\ \emph {et~al.}(2015)\citenamefont {Mori},
  \citenamefont {Kuwahara},\ and\ \citenamefont {Saito}}]{Mori2015}%
  \BibitemOpen
  \bibfield  {author} {\bibinfo {author} {\bibfnamefont {Takashi}\ \bibnamefont
  {Mori}}, \bibinfo {author} {\bibfnamefont {Tomotaka}\ \bibnamefont
  {Kuwahara}}, \ and\ \bibinfo {author} {\bibfnamefont {Keiji}\ \bibnamefont
  {Saito}},\ }\bibfield  {title} {\enquote {\bibinfo {title} {Rigorous bound on
  energy absorption and generic relaxation in periodically driven quantum
  systems},}\ }\href@noop {} {\bibfield  {journal} {\bibinfo  {journal}
  {arXiv:1509.03968v2}\ } (\bibinfo {year} {2015})}\BibitemShut {NoStop}%
\bibitem [{\citenamefont {Thouless}(1983)}]{Thouless1983}%
  \BibitemOpen
  \bibfield  {author} {\bibinfo {author} {\bibfnamefont {D.~J.}\ \bibnamefont
  {Thouless}},\ }\bibfield  {title} {\enquote {\bibinfo {title} {Quantization
  of particle transport},}\ }\href@noop {} {\bibfield  {journal} {\bibinfo
  {journal} {Phys. Rev. B}\ }\textbf {\bibinfo {volume} {27}},\ \bibinfo
  {pages} {6083--6087} (\bibinfo {year} {1983})}\BibitemShut {NoStop}%
\bibitem [{\citenamefont {Lohse}\ \emph {et~al.}(2015)\citenamefont {Lohse},
  \citenamefont {Schweizer}, \citenamefont {Zilberberg}, \citenamefont
  {Aidelsburger},\ and\ \citenamefont {Bloch}}]{lohse2015thouless}%
  \BibitemOpen
  \bibfield  {author} {\bibinfo {author} {\bibfnamefont {Michael}\ \bibnamefont
  {Lohse}}, \bibinfo {author} {\bibfnamefont {Christian}\ \bibnamefont
  {Schweizer}}, \bibinfo {author} {\bibfnamefont {Oded}\ \bibnamefont
  {Zilberberg}}, \bibinfo {author} {\bibfnamefont {Monika}\ \bibnamefont
  {Aidelsburger}}, \ and\ \bibinfo {author} {\bibfnamefont {Immanuel}\
  \bibnamefont {Bloch}},\ }\bibfield  {title} {\enquote {\bibinfo {title} {A
  thouless quantum pump with ultracold bosonic atoms in an optical
  superlattice},}\ }\href@noop {} {\bibfield  {journal} {\bibinfo  {journal}
  {arXiv:1507.02225; Nature Physics, in press}\ } (\bibinfo {year}
  {2015})}\BibitemShut {NoStop}%
\bibitem [{\citenamefont {Nakajima}\ \emph {et~al.}(2016)\citenamefont
  {Nakajima}, \citenamefont {Tomita}, \citenamefont {Taie}, \citenamefont
  {Ichinose}, \citenamefont {Ozawa}, \citenamefont {Wang}, \citenamefont
  {Troyer},\ and\ \citenamefont {Takahashi}}]{nakajima2016topological}%
  \BibitemOpen
  \bibfield  {author} {\bibinfo {author} {\bibfnamefont {Shuta}\ \bibnamefont
  {Nakajima}}, \bibinfo {author} {\bibfnamefont {Takafumi}\ \bibnamefont
  {Tomita}}, \bibinfo {author} {\bibfnamefont {Shintaro}\ \bibnamefont {Taie}},
  \bibinfo {author} {\bibfnamefont {Tomohiro}\ \bibnamefont {Ichinose}},
  \bibinfo {author} {\bibfnamefont {Hideki}\ \bibnamefont {Ozawa}}, \bibinfo
  {author} {\bibfnamefont {Lei}\ \bibnamefont {Wang}}, \bibinfo {author}
  {\bibfnamefont {Matthias}\ \bibnamefont {Troyer}}, \ and\ \bibinfo {author}
  {\bibfnamefont {Yoshiro}\ \bibnamefont {Takahashi}},\ }\bibfield  {title}
  {\enquote {\bibinfo {title} {Topological thouless pumping of ultracold
  fermions},}\ }\href@noop {} {\bibfield  {journal} {\bibinfo  {journal}
  {arXiv:1507.02223; Nature Physics, in press}\ } (\bibinfo {year}
  {2016})}\BibitemShut {NoStop}%
\bibitem [{Note1()}]{Note1}%
  \BibitemOpen
  \bibinfo {note} {For any finite drive frequency $\omega $, minigaps open at
  the crossing between the Floquet bands. These minigaps are suppressed
  exponentially in $1/\omega $.}\BibitemShut {Stop}%
\bibitem [{Note2()}]{Note2}%
  \BibitemOpen
  \bibinfo {note} {A conservative estimate of the corrections to this value of
  the current is $\protect \mathrm {O}[(U/\Delta )^2]$, due to virtual
  transitions between the bands.}\BibitemShut {Stop}%
\bibitem [{\citenamefont {Deutsch}(1991)}]{Deutsch91}%
  \BibitemOpen
  \bibfield  {author} {\bibinfo {author} {\bibfnamefont {J.~M.}\ \bibnamefont
  {Deutsch}},\ }\bibfield  {title} {\enquote {\bibinfo {title} {Quantum
  statistical mechanics in a closed system},}\ }\href@noop {} {\bibfield
  {journal} {\bibinfo  {journal} {Phys. Rev. A}\ }\textbf {\bibinfo {volume}
  {43}},\ \bibinfo {pages} {2036} (\bibinfo {year} {1991})}\BibitemShut
  {NoStop}%
\bibitem [{\citenamefont {Srednicki}(1999)}]{Srednicki99}%
  \BibitemOpen
  \bibfield  {author} {\bibinfo {author} {\bibfnamefont {M.}~\bibnamefont
  {Srednicki}},\ }\bibfield  {title} {\enquote {\bibinfo {title} {The approach
  to thermal equilibrium in quantized chaotic systems},}\ }\href@noop {}
  {\bibfield  {journal} {\bibinfo  {journal} {J. Phys. A: Math. Gen.}\ }\textbf
  {\bibinfo {volume} {32}},\ \bibinfo {pages} {1163} (\bibinfo {year}
  {1999})}\BibitemShut {NoStop}%
\bibitem [{\citenamefont {Reimann}(2008)}]{Reimann2008}%
  \BibitemOpen
  \bibfield  {author} {\bibinfo {author} {\bibfnamefont {P.}~\bibnamefont
  {Reimann}},\ }\bibfield  {title} {\enquote {\bibinfo {title} {Foundation of
  statistical mechanics under experimentally realistic conditions},}\
  }\href@noop {} {\bibfield  {journal} {\bibinfo  {journal} {Phys. Rev. Lett.}\
  }\textbf {\bibinfo {volume} {101}},\ \bibinfo {pages} {190403} (\bibinfo
  {year} {2008})}\BibitemShut {NoStop}%
\bibitem [{Note3()}]{Note3}%
  \BibitemOpen
  \bibinfo {note} {We take $\hbar = 1$ throughout.}\BibitemShut {Stop}%
\bibitem [{\citenamefont {Choudhury}\ and\ \citenamefont
  {Mueller}(2015)}]{Choudhury2015}%
  \BibitemOpen
  \bibfield  {author} {\bibinfo {author} {\bibfnamefont {Sayan}\ \bibnamefont
  {Choudhury}}\ and\ \bibinfo {author} {\bibfnamefont {Erich~J.}\ \bibnamefont
  {Mueller}},\ }\bibfield  {title} {\enquote {\bibinfo {title} {Transverse
  collisional instabilities of a bose-einstein condensate in a driven
  one-dimensional lattice},}\ }\href {\doibase 10.1103/PhysRevA.91.023624}
  {\bibfield  {journal} {\bibinfo  {journal} {Phys. Rev. A}\ }\textbf {\bibinfo
  {volume} {91}},\ \bibinfo {pages} {023624} (\bibinfo {year}
  {2015})}\BibitemShut {NoStop}%
\bibitem [{\citenamefont {Bilitewski}\ and\ \citenamefont
  {Cooper}(2015)}]{Bilitewski2015}%
  \BibitemOpen
  \bibfield  {author} {\bibinfo {author} {\bibfnamefont {Thomas}\ \bibnamefont
  {Bilitewski}}\ and\ \bibinfo {author} {\bibfnamefont {Nigel~R}\ \bibnamefont
  {Cooper}},\ }\bibfield  {title} {\enquote {\bibinfo {title} {Scattering
  theory for floquet-bloch states},}\ }\href@noop {} {\bibfield  {journal}
  {\bibinfo  {journal} {Physical Review A}\ }\textbf {\bibinfo {volume} {91}},\
  \bibinfo {pages} {033601} (\bibinfo {year} {2015})}\BibitemShut {NoStop}%
\bibitem [{\citenamefont {Genske}\ and\ \citenamefont
  {Rosch}(2015)}]{Genske2015}%
  \BibitemOpen
  \bibfield  {author} {\bibinfo {author} {\bibfnamefont {Maximilian}\
  \bibnamefont {Genske}}\ and\ \bibinfo {author} {\bibfnamefont {Achim}\
  \bibnamefont {Rosch}},\ }\bibfield  {title} {\enquote {\bibinfo {title}
  {Floquet-boltzmann equation for periodically driven fermi systems},}\ }\href
  {\doibase 10.1103/PhysRevA.92.062108} {\bibfield  {journal} {\bibinfo
  {journal} {Phys. Rev. A}\ }\textbf {\bibinfo {volume} {92}},\ \bibinfo
  {pages} {062108} (\bibinfo {year} {2015})}\BibitemShut {NoStop}%
\bibitem [{Note4()}]{Note4}%
  \BibitemOpen
  \bibinfo {note} {{The Fermi's golden rule rates are evaluated using ``free''
  initial and final states, as taken at $t = 0$. For $t \gtrsim \tau _{\protect
  \rm intra}$, the state locally appears similar to an infinite temperature
  state (within the R band), which can be described as a uniform mixture of all
  free states in the band.}}\BibitemShut {Stop}%
\bibitem [{com()}]{comment-quasienergy}%
  \BibitemOpen
  \href@noop {} {}\bibinfo {note} {We use a convention where the quasi-energies
  lie in the window $[0,\omega)$.}\BibitemShut {Stop}%
\bibitem [{dou()}]{doublon}%
  \BibitemOpen
  \href@noop {} {}\bibinfo {note} {The consideration here are similar to the
  ones employed in the problem of doublon decay in a Mott insulator. E.g., see:
  R, Sensarma, D. Pekker, M. D. Lukin, E. Demler, Phys. Rev. Lett.
  \textbf{103}, 035303 (2009); Z. Lenar{\v{c}}i{\v{c}}, M. Eckstein, and P.
  Prelov{\v{s}}ek, Phys. Rev. B \textbf{92}, 201104 (2015).}\BibitemShut
  {Stop}%
\bibitem [{\citenamefont {Aidelsburger}\ \emph {et~al.}(2015)\citenamefont
  {Aidelsburger}, \citenamefont {Lohse}, \citenamefont {Schweizer},
  \citenamefont {Atala}, \citenamefont {Barreiro}, \citenamefont
  {Nascimb{\`e}ne}, \citenamefont {Cooper}, \citenamefont {Bloch},\ and\
  \citenamefont {Goldman}}]{aidelsburger2015measuring}%
  \BibitemOpen
  \bibfield  {author} {\bibinfo {author} {\bibfnamefont {Monika}\ \bibnamefont
  {Aidelsburger}}, \bibinfo {author} {\bibfnamefont {Michael}\ \bibnamefont
  {Lohse}}, \bibinfo {author} {\bibfnamefont {C}~\bibnamefont {Schweizer}},
  \bibinfo {author} {\bibfnamefont {Marcos}\ \bibnamefont {Atala}}, \bibinfo
  {author} {\bibfnamefont {Julio~T}\ \bibnamefont {Barreiro}}, \bibinfo
  {author} {\bibfnamefont {S}~\bibnamefont {Nascimb{\`e}ne}}, \bibinfo {author}
  {\bibfnamefont {NR}~\bibnamefont {Cooper}}, \bibinfo {author} {\bibfnamefont
  {Immanuel}\ \bibnamefont {Bloch}}, \ and\ \bibinfo {author} {\bibfnamefont
  {N}~\bibnamefont {Goldman}},\ }\bibfield  {title} {\enquote {\bibinfo {title}
  {{Measuring the Chern number of Hofstadter bands with ultracold bosonic
  atoms}},}\ }\href@noop {} {\bibfield  {journal} {\bibinfo  {journal} {Nature
  Physics}\ }\textbf {\bibinfo {volume} {11}},\ \bibinfo {pages} {162--166}
  (\bibinfo {year} {2015})}\BibitemShut {NoStop}%
\bibitem [{Note5()}]{Note5}%
  \BibitemOpen
  \bibinfo {note} {This is important since, at this point, the Floquet
  eigenstates are mixtures of states from the upper and lower instantaneous
  bands; we would like to initialize the system with states that are mostly
  from the instantaneous lower energy band.}\BibitemShut {Stop}%
\end{thebibliography}%

\end{document}